\documentclass[10pt]{article}

\usepackage[a4paper,margin=1in]{geometry}
\usepackage{amsmath, amssymb, amsfonts}
\usepackage{graphicx}
\usepackage{hyperref}
\usepackage{xcolor}
\usepackage{cite}
\usepackage{booktabs}
\usepackage{physics}
\usepackage{subcaption} 

\newcommand{\nband}{n_{\mathrm{b}}}

\title{Scalable topological quantum computing with Matryoshka-type sine--cosine chains}

\author{
	A. Lykholat$^{1,4}$,
	G. F. Moreira$^1$,
	I. R. Martins$^2$, 
	D. Sousa$^3$,\\
	A. M. Marques$^1$, 
	R. G. Dias$^{1*}$ \\
	\\
	$^1$Department of Physics \& i3N, University of Aveiro, \\3810-193 Aveiro, Portugal \\
	$^2$Laboratório de Instrumentação e Física Experimental de Partículas (LIP), \\
	Universidade do Minho, 4710-057 Braga, Portugal \\
	$^3$Max-Planck-Institute for Solid State Research, Heisenbergstrasse 1, \\ 70569 Stuttgart, Germany \\
	$^4$Department of Theoretical Physics, Institute of Physics, \\ Budapest University of Technology and Economics, \\ Műegyetem rkp. 3., H-1111 Budapest, Hungary \\
	$^*$rdias@ua.pt
}

\date{}

\begin{document}
	
	\maketitle
	
\begin{abstract}
We propose a scalable framework for quantum information processing based on Matryoshka-type sine--cosine chains, a hierarchy of one-dimensional chiral-symmetric lattices obtained by repeated ($P$ times) square-root construction starting from the Su--Schrieffer--Heeger (SSH) chain. A chain of order $P$ can host $2^{P+2}-2$ protected edge states, so a single physical device can encode a qudit of equivalent dimension  (or, equivalently, several independent qubits) instead of one qubit per device. We show that (i) the adiabatic defect-transfer step underlying SSH-based  topological gates carries over unchanged to the Matryoshka chain, transferring the full set of defect states simultaneously; (ii) a Y-junction braiding protocol then composes single-qudit gate operations in all energy sectors at once, increasing the operation density per physical device; and (iii) the same edge states provide a quantum memory whose capacity scales exponentially with the chain order at fixed physical length. Numerical fidelity analysis under on-site, hopping, and correlated angle disorder shows partial, chiral-symmetry-protected robustness.
\end{abstract}


\medskip

\section{Introduction}

Practical quantum computing is limited by the fragility of quantum states: environmental interactions cause decoherence and information loss \cite{Zurek2003,Shor1995}. Despite decades of progress, dominant architectures, such as superconducting qubits, \cite{Clarke2008, Gibney2014}, struggle with scalability. The physical resource overhead of fault tolerance (building each logical qubit from many physical qubits and running error correction continuously) grows rapidly with system size \cite{Fowler2012, Shor1996}. These challenges motivate approaches in which information is protected by the topology of the physical system rather than by error correction alone \cite{Kitaev2003,Nayak2008}.

Square-root topological insulators ($\sqrt{\text{TI}}$) \cite{Ezawa2020,Arkinstall2017} form a class of systems whose topological features are hidden until the Hamiltonian is squared, revealing a ``parent'' topological insulator \cite{Kremer2020,Pelegri2019,Mizoguchi2020,Ezawa2020,Ke2020,Mizoguchi2021c,Yoshida2021,Lin2021,Wu2021,Bomantara2022,Zhang2022,Matsumoto2023,Geng2023,Roy2024,Zhao2025}. Because of this hidden relationship, the edge states of a $\sqrt{\text{TI}}$ retain part of the protection enjoyed by the parent's zero-energy edge states. The logic generalizes to $2^n$-root topological insulators, obtained by stacking $n$ square-root operations (with accompanying energy shifts) \cite{Marques2021,Marques2021a,Dias2021,Moreira2026,Dias2025,Marques2023,Geng2024}; the resulting recursive structure can be visualized as an arborescence \cite{Marques2021a} or, for the specific sine-cosine models considered here, as a Matryoshka sequence \cite{Dias2021,Dias2025}. As the root degree grows, the number of edge states increases, but the protection inherited from each parent level is progressively diluted. Encoding several qubits or a single high-dimensional qudit within one such system therefore reduces the number of independent physical devices required per logical degree of freedom while retaining partial robustness against  perturbations.

This paper studies a particular member of this family: the Matryoshka model \cite{Dias2021}, obtained by recursively applying  square-root operations to the Su-Schrieffer-Heeger (SSH) chain \cite{Su1979}. The construction produces a lattice with staggered sine-cosine hopping terms, a larger number of energy gaps, and a larger number of edge states. The main contributions of this paper are the following:
(i) we give a compact, self-contained description of the Matryoshka chain of order $P$ (Sec.~\ref{sec:model}), including its band structure, its protected edge states, and the role of chiral symmetry;
(ii) we show that the adiabatic defect-transfer step used in the poor-man's topological gate of Boross \emph{et al.} \cite{Boross2019} carries over unchanged to the Matryoshka chain, and that the \emph{full set} of defect states (one per protected gap) can be transferred simultaneously (Sec.~\ref{sec:transfer});
(iii) we build a Y-junction braiding protocol for the Matryoshka chain and show that a single braiding event applies single-qudit gate operations in all energy sectors at once, increasing the operation density per device (Sec.~\ref{sec:gates}). We quantify robustness against on-site, hopping, and correlated angle disorder;
(iv) we embed the Matryoshka chain in the SSH quantum-memory protocol \cite{palyi2018ssh} and show that the memory capacity scales exponentially with the order $P$ at fixed physical length (Sec.~\ref{sec:memory}, App.~\ref{app:qudit}).

The remainder of the paper is organized as follows. Section \ref{sec:model} defines the Matryoshka chain. Section \ref{sec:transfer} describes adiabatic defect transfer in these chains. Section \ref{sec:gates} presents the Y-junction braiding gate and its robustness to disorder. Section \ref{sec:memory} extends the analysis to quantum memory architectures. Section \ref{sec:conclusion} summarizes the scalability benefits, states the limitations of the present approach, and discusses experimental realizations. Appendices \ref{app:transfer}--\ref{app:qudit} provide the transfer-protocol derivation, the level-crossing and disorder analysis, and the multi-qubit/qudit memory extension, respectively.

\section{The Matryoshka Model}\label{sec:model}

\begin{figure}[ht!]
  \centering  
  \includegraphics[
width=0.9\textwidth]{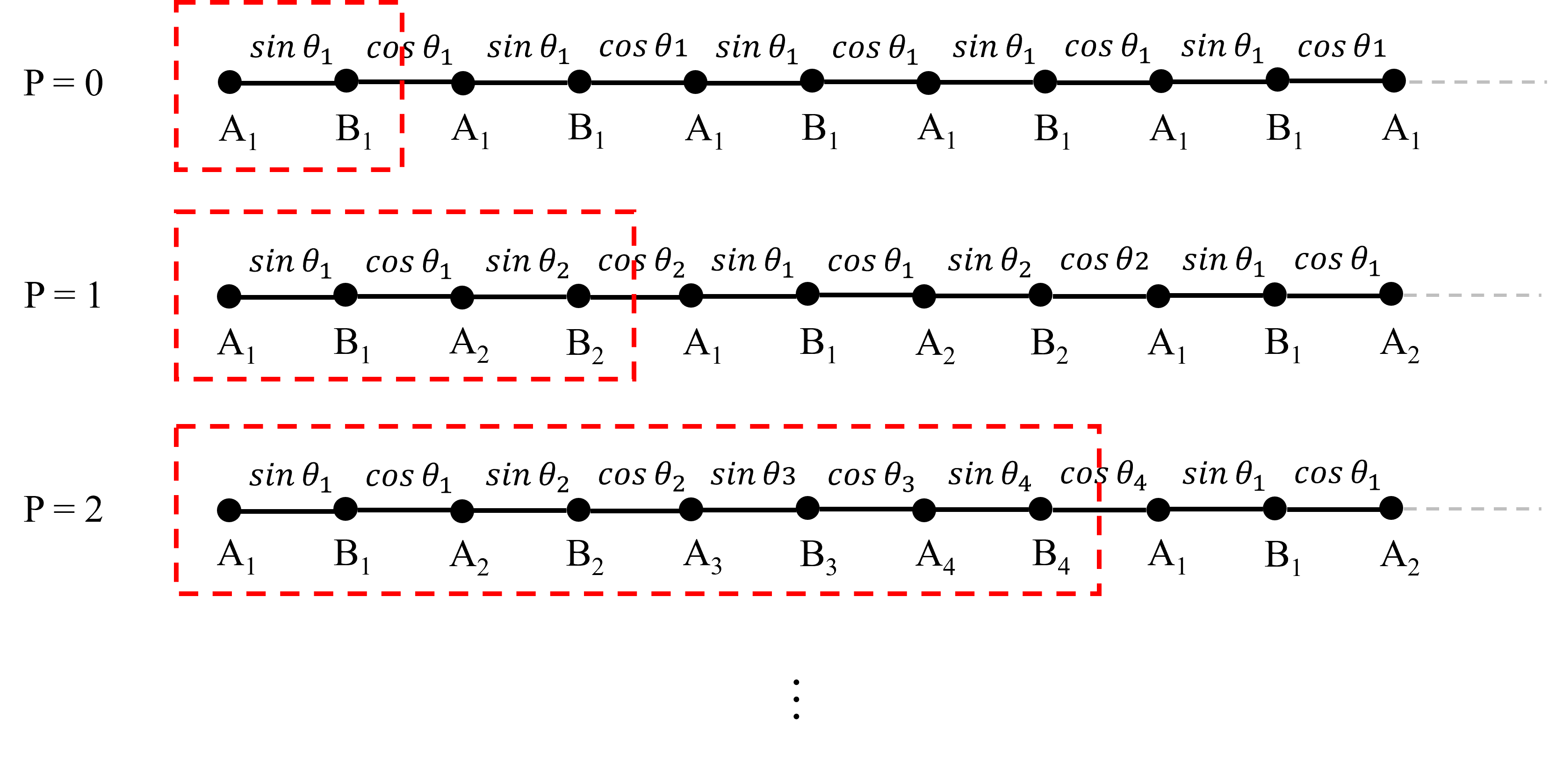}
  \caption{Lattice representation of the first orders of $P$ in the Matryoshka model. The figure illustrates how the unit cell (dashed red box) evolves for $P = 0, 1, 2$, becoming progressively larger as we increase $P$, with hopping terms given by sine and cosine functions of angles $\theta_j$.}
  \label{fig:matryoska_base}
\end{figure}

In this section we introduce the Matryoshka model \cite{Dias2021}, which supports significantly more edge states than the SSH chain and, as we will show, can host multiple qubits or a high-dimensional qudit within a single physical device. We use the notation $H^{(P)}$ for the chain of order $P$ (so $H^{(0)}$ is the SSH chain). A Matryoshka chain is composed of a sequence of sine and cosine hopping terms,
\begin{equation}
	\begin{aligned}
	H^{(P)} = \; &  t^{(P)} \sum_{m=1}^N \Bigg[  \sum_{j=1}^{2^P} \sin{\theta_j} \ket{m, B_{j}}\bra{m, A_j} + \sum_{j=1}^{2^P-1} \cos{\theta_j} \ket{m, A_{j+1}}\bra{m, B_j}  \\
	& +  \cos{\theta_{2^{P}}} \ket{m+1, A_{1}}\bra{m, B_{2^P}} \Bigg] + H.c. ,
	\end{aligned}
\end{equation}
where $P$ is the structure order, $N$ is the number of unit cells, and $t^{(P)}$ is the overall energy scale. The number of distinct hopping angles is $2^P$, growing exponentially with $P$; Fig.~\ref{fig:matryoska_base} shows the unit cell for the first three orders.

\begin{figure}[ht!]
\centering  
  \includegraphics[
width=0.8\textwidth]{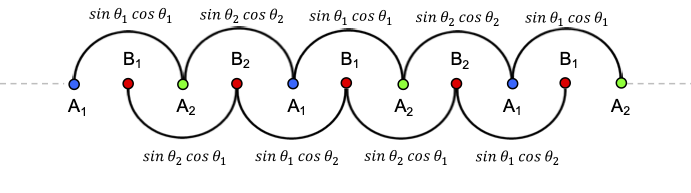}
  \caption{Representation of the resulting structures $H_A$ and $H_B$ after applying the squaring operation to the parent Hamiltonian. The diagram illustrates the separation into two decoupled sublattices, with effective hoppings dependent on the parameters $\theta_1$ and $\theta_2$. Different site colors represent different onsite potentials.}
  \label{fig:splitting}
\end{figure}

The $P=0$ system, with the two connection values $\sin\theta_1$ and $\cos\theta_1$, is exactly the SSH chain, gapped at the center of the spectrum for $\theta_1\neq \pi/4$. In the edge-dimerized limit $\sin\theta_1=1$ (strong end-dimer) and $\cos\theta_1=0$ (decoupled cells) it reduces to a chain of isolated dimers.
The Matryoshka Hamiltonian is bipartite, so when squared in the chiral basis $\{\{\ket{m,A_j}\},\{\ket{m,B_j}\}\}$, with $j=1,2,\dots,2^P$ and $m=1,2,\dots,N$, yields, for general $P$,
\begin{equation}
\left(\dfrac{H^{(P)}}{t^{(P)} } \right)^2= \begin{pmatrix} H_{A} & 0 \\ 0 & H_{B} \end{pmatrix}.
\label{eq:squaring}
\end{equation}
In the case $P=1$, $H_B$ is a bipartite structure equivalent to the $P=0$ chain with a global energy offset of $+1$ (red sites in Fig.~\ref{fig:splitting}), while $H_A$ is a residual chain with the same finite energy spectrum \cite{Ezawa2020} but with staggered on-site potentials along its matrix diagonal (blue and green sites in Fig.~\ref{fig:splitting}).
Because the $H_B$ block is the parent chain (order $P-1$) shifted in energy by $+1$, the spectra of $H^{(P-1)}$ and $H_B$ coincide apart from that shift. For general $P$ this imposes the following relations on the hopping angles:
\begin{equation}
t^{(P-1)} \sin{\theta_j}^{(P-1)} = \sin{\theta_{2j-1}}^{(P)} \cos{\theta_{2j}}^{(P)},
\end{equation}
\begin{equation}
t^{(P-1)} \cos{\theta_j}^{(P-1)} = \sin{\theta_{2j}}^{(P)} \cos{\theta_{2j+1}}^{(P)},
\end{equation}
\begin{equation}
t^{(P-1)} = \sqrt{(\sin{\theta_{2j-1}}^{(P)}\cos{\theta_{2j}}^{(P)})^2 + (\sin{\theta_{2j}}^{(P)}\cos{\theta_{2j+1}}^{(P)})^2},
\end{equation}
with $j=1,\dots,2^{P-1}$.

The $P=0$ energy dispersion (of $H^{(0)}/t^{(0)}$) is $E_{\pm}^{(0)}(k) = \pm\sqrt{1 + \sin{2\theta}\cos{k}}$, and for general $P$ we obtain the multi-band energy spectrum (of $H^{(P)}/t^{(P)}$)
\begin{equation}
E^{(P)} _{\pm\pm\cdots\pm}(k)= \pm \sqrt{1 + t^{(P-1)}E^{(P-1)}_{\pm\pm\cdots\pm}(k)}.
\label{eq:bands_rec}
\end{equation}
Each band label carries $P+1$ signs, and all sign combinations must be considered. Equation \eqref{eq:bands_rec} shows that the number of energy bands doubles at each squaring step, $\nband^{(P)} = 2\,\nband^{(P-1)} = 2^{P+1}$, with an energy offset of $1$ accumulated after each squaring operation.
\begin{figure}[ht!]
 \centering  
 \includegraphics[
width=0.9\textwidth]{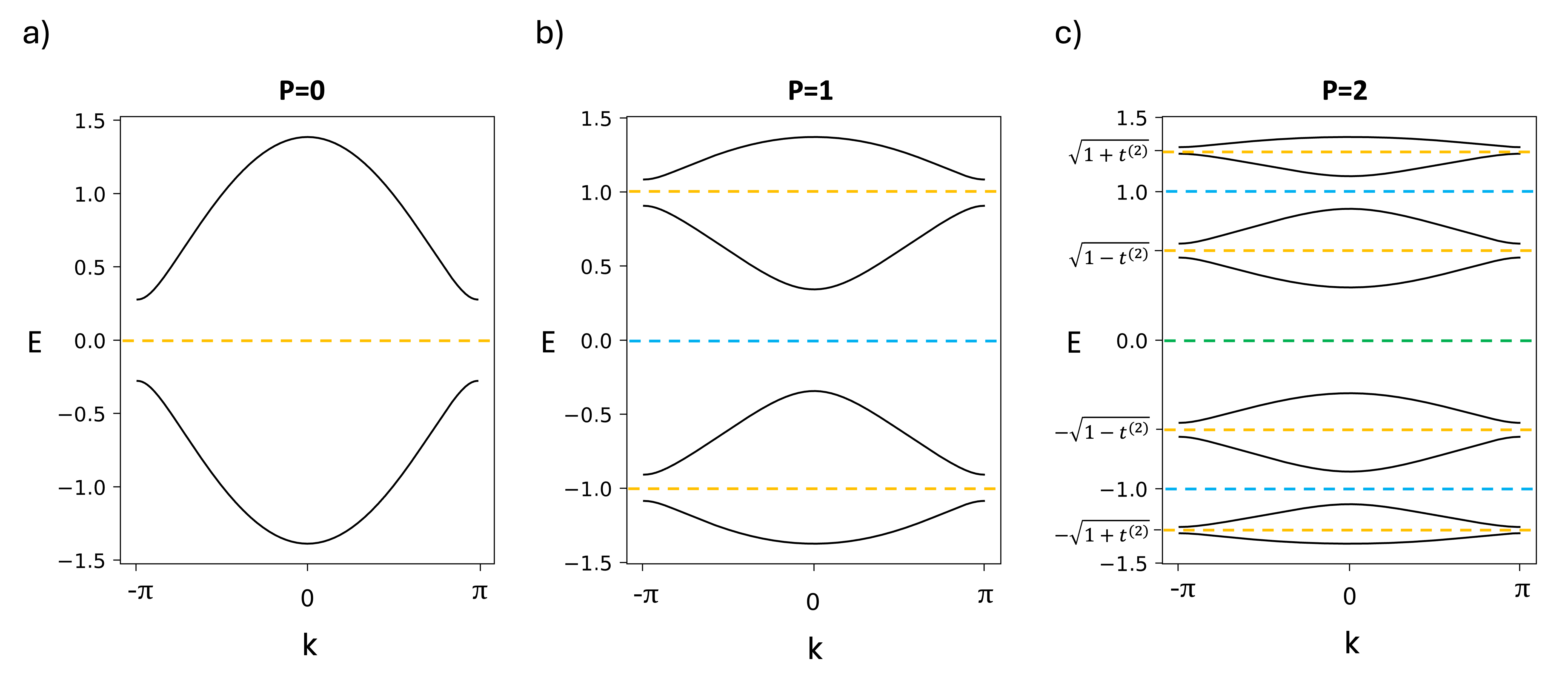}
  \caption{Band structure of (a) $P=0$, (b) $P=1$ and (c) $P=2$ Matryoshka chain. To calculate the base lattice ($P=0$) energies, $\sin{\theta_1}=0.588$ was used. To obtain the next lattice's band structure $t^{(0)}=0.9/\sqrt{2}$ and $t^{(1)}=0.8/\sqrt{2}$ were used. The yellow line represents the edge states relative to the weak links of $P=0$, which double after each squaring. In the case that $P=0$ does not present weak links, none of the yellow lines can be seen in the spectra. This same logic applies to the degenerate blue states in $P=1$, which unfold and form two in $P=2$. The green line corresponds to the weak links for $P=2$.}
  \label{fig:bands}
\end{figure}
All energy spectra are chiral-symmetric, $\Gamma H^{(P)} (k) \Gamma^{-1}= - H^{(P)} (k)$ where $\Gamma =\sigma_z \otimes \mathbb{I}_{2^P} $ with $\sigma_z$ the third Pauli matrix and  $\mathbb{I}_j $ the identity matrix of size $j$. This is evident in Fig.~\ref{fig:bands}.

The existence of weak links at the chain ends in any structure of type $P$ implies that two degenerate (ignoring finite size corrections) edge states of energy $E=0$ exist in its spectrum. 
Each zero-energy level of order $P$ unfolds into two levels in the spectrum of order $P+1$. Correlated disorder applied to the angles $\theta_j$ of the chain of order $P+1$  does not break the chiral symmetry of the parent chain of order  $P$  \cite{Moreira2026}, so the protection of the unfolded edge states  is preserved. The energies at which protected edge states can appear are the nested square-root values

\begin{align}
& \pm 1, \quad \pm \sqrt{1 \pm t^{(P-1)}}, \quad \pm \sqrt{1 \pm t^{(P-1)} \sqrt{1 \pm t^{(P-2)}}}, \notag \\ 
& \cdots \quad \pm \sqrt{1 \pm t^{(P-1)} \sqrt{1 \pm t^{(P-2)} \sqrt{\,\cdots\, \sqrt{1 \pm t^{(1)}}}}}.
\label{eq:edge_energies}
\end{align}

In summary, increasing the order $P$ doubles the number of energy bands, to $\nband^{(P)}=2^{P+1}$, and hence the total number of inter-band gaps, $2^{P+1}-1$.  Under open boundary conditions (OBC) a chain of order $P$ therefore supports up to
\begin{equation}
N_{\mathrm{edge}}^{(P)} = 2\left(2^{P}-1\right) = 2^{P+2}-2
\label{eq:n_edge}
\end{equation}
protected edge states (two zero-energy states  plus the unfolded parent edge states). For $P=0$ this gives the two SSH edge states; for $P=1$ it gives six (the two $\varepsilon=0$ states plus the four $\varepsilon=\pm1$ states); for $P=2$ it gives fourteen.

\section{Adiabatic defect transfer in Matryoshka chains}\label{sec:transfer}

Quantum state transfer along one-dimensional (1D) chains has attracted considerable attention for scalable quantum information processing and quantum communication \cite{dAngelis2020,Dlaska2017,Mei2018,Zhang2024,Zurita2023}. In 1D topological models such as the SSH chain, the transport of a particle from one end of the chain to the other can be mediated by protected edge states, which provide robustness against disorder and environmental noise \cite{Mei2018, Zhang2024,Estarellas2017}.

\begin{figure}[ht!]
 \centering  
 \includegraphics[
width=0.9\textwidth]{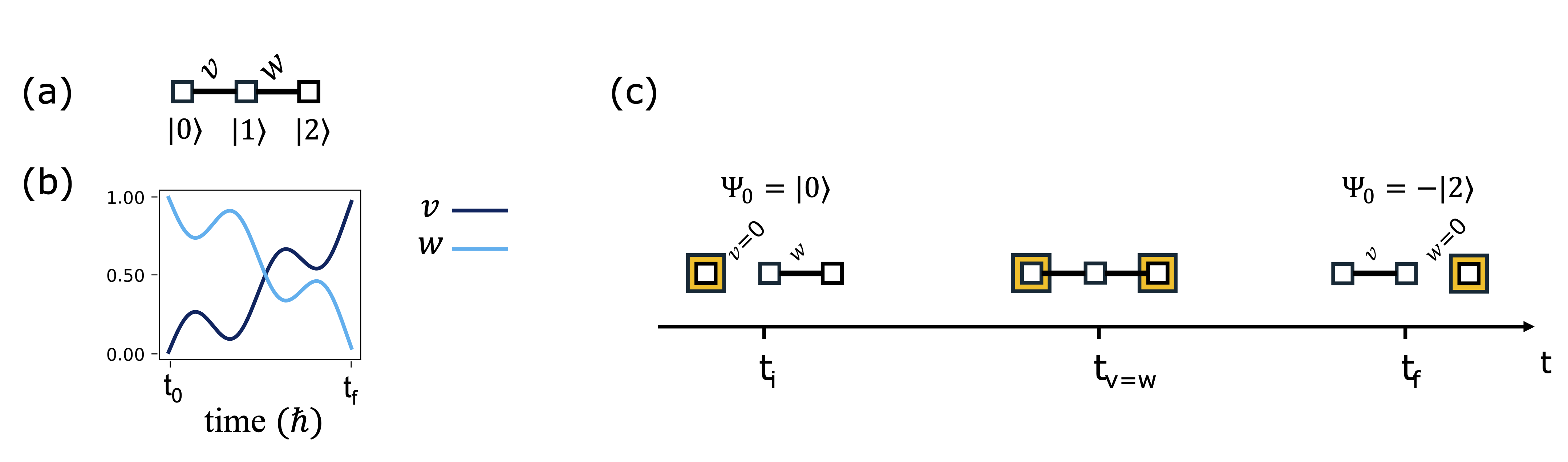}
  \caption{In this figure, we reproduce the transferring protocol of \cite{Boross2019}. (a) Base system. (b) Adiabatic sweep of the couplings $v(t)$ and $w(t)$. (c) Transfer protocol illustration. The yellow squares indicate finite state amplitudes. When $v = 0$ we have $\ket{\Psi(t_i)} = \ket{0}$, and $\ket{\Psi (t_f)} = -\ket{2}$ if $w = 0$. For the case where $v = w$, the state will be equally distributed between $\ket{0}$ and $\ket{2}$.}
  \label{fig:transfer_protocol}
\end{figure}

A key ingredient of the poor-man's topological gate of Boross \emph{et al.} \cite{Boross2019} is the \emph{adiabatic motion of a defect} (a localized zero-energy mode) along a short SSH segment, as illustrated in Fig.~\ref{fig:transfer_protocol}. In this section we show that this defect-transfer step carries over unchanged to the Matryoshka chain, because the square-root construction implies that the eigenvectors of the Matryoshka chain are built from those of the parent SSH chain. We first review the three-site parent system, then state the parent/square-root eigenvector mapping, and finally give the angle-schedule protocol that transfers the \emph{entire} set of Matryoshka defect states simultaneously.

The three-site SSH chain in Fig.~\ref{fig:transfer_protocol}(a) is a bipartite tight‑binding small chain with an imbalance in the number of sites on the sublattices, which implies the existence of a zero‑energy eigenstate. The diagonalization of this system shows that the normalized zero‑energy state is given by
\begin{equation}
\ket{\Psi_0} = \frac{w\ket{0}-v\ket{2}}{\sqrt{v^2+w^2}}.
\end{equation}
The initially decoupled site ($v=0$ case) is designated a \emph{defect} (or domain wall); if more dimers are present in the chain, the defect can be transferred along it. As discussed in Sec.~\ref{sec:gates}, quantum gates are constructed by braiding this defect in a Y-junction configuration. The essential point is that, if the transfer is slow, adiabaticity is preserved: the defect state is, in fact, a zero-energy edge state of an SSH chain with a non-integer number of unit cells. An evolution in which $v$ increases from 0 to 1, as shown in Fig.~\ref{fig:transfer_protocol}(b), while $w$ decreases from 1 to 0 then transfers the defect (with a $\pi$ phase shift), as shown in Fig.~\ref{fig:transfer_protocol}(c), without any energy-level crossing (the gap between the defect level and the rest of the spectrum is $\sqrt{v^2 + w^2}$, finite as long as $v$ and $w$ are not simultaneously zero).

\begin{figure}[ht!]
  \includegraphics[
width=0.9\textwidth]{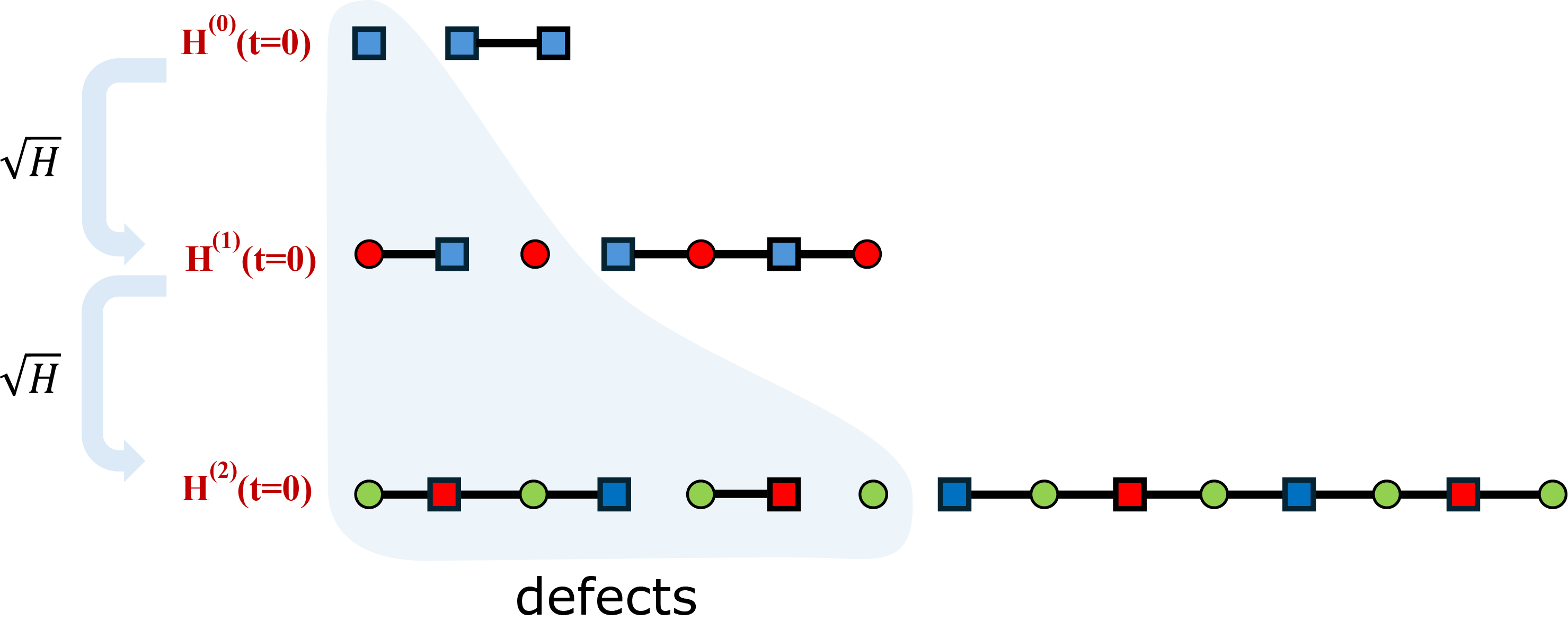}
  \caption{Consecutive square root of the initial Hamiltonian in the quantum transfer process. The highlighted defect state ($H^{(1)}$) corresponds to the parent Hamiltonian's full chain  ($H^{(0)}$), where single-site, dimer, and 4-site cluster configurations are interchangeable under adiabatic evolution. The highlight shows the defect part in each model.  $H^{(1)}(t=0)$ corresponds to $\theta_{1,i} = \pi/2$, $\theta_{2,i} = 0$, and $\theta_{3,i} = \lambda$. In the first chain, we have one angle, in the second, we have three, and in the third, we have seven (see Appendix A for details). The amplitudes of the original three-site SSH chain eigenstates will be present in the blue sites of all chains, except for eigenstates that result from the zero energy eigenstates of $H_{(1)}$, $H_{(2)}$, etc.}
  \label{fig:consecutive}
\end{figure}

We now extend the same logic to the sine--cosine chain of Sec.~\ref{sec:model}. The square-root construction that turns the parent Hamiltonian of an SSH chain into its square-root sine--cosine counterpart does not alter the underlying eigenstates, except for an overall phase or normalization factor. Let the parent Hamiltonian (SSH chain) be $H^{(0)}$ and its square-root counterpart be $H^{(1)}$. The key relation is
\begin{equation}
\left(\dfrac{H^{(1)}}{t^{(1)} } \right)^2=
\begin{pmatrix}
H_A & 0 \\
0   & H_B
\end{pmatrix},
\end{equation}
where $H_B$ is the parent-chain Hamiltonian shifted in energy by $+1$, $H_B=H^{(0)}+\mathbb{I}$. An eigenstate $\ket{\psi_E}$ of $H^{(1)}$ can be written as $\ket{\psi_E} = \alpha\,\ket{\phi_A}+ \beta\,\ket{\phi_B}$, with amplitudes $\alpha$ and $\beta$ on the two decoupled sublattices. Because $H^{(0)}$ has a protected zero-energy mode $\ket{\Psi_0}$, the corresponding defect state of the bipartite $H^{(1)}/t^{(1)} $ has energy $\varepsilon=\pm1$ and is
\begin{equation}
\ket{\psi_{\varepsilon=\pm1}} = 
\frac{1}{\sqrt{2}}
\Bigl(
\ket{\Psi_{\text{A}}} \pm\ket{\Psi_0}
\Bigr),
\end{equation}
where $\ket{\Psi_{\text{A}}} = H^{(1)}/t^{(1)}  \ket{\Psi_0}$.
Thus the parent defect eigenstate maps onto one sublattice component of a square-root-chain defect eigenstate, up to a phase. The argument iterates to higher roots: each squaring step doubles the defect space and dilutes the weight carried by the original SSH sublattice by a factor $1/\sqrt{2}$, as illustrated in Fig.~\ref{fig:consecutive}.

During a quasi-adiabatic evolution, in which the hopping angles $\theta_j(t)$ change slowly, the adiabatic theorem guarantees that a system prepared in an eigenstate of the parent Hamiltonian remains in that instantaneous eigenstate provided the energy levels do not cross (details in Appendix \ref{app:robustness}), accumulating only dynamical and Berry phases. Because an eigenvector of the parent Hamiltonian (with energy $E^2-1$, where $E$ is the square-root-chain energy) coincides with a sublattice component of an eigenvector of $H^{(1)}$ (with energy $E$), the adiabatic trajectory of the parent is inherited by the square-root system. The zero-energy defect of the SSH parent maps onto the $\varepsilon=\pm1$ defect pair of the Matryoshka chain (Fig.~\ref{fig:consecutive}), so the transfer protocol works unchanged up to the additional phase that naturally accumulates during the evolution. $H^{(1)}$ remains in the corresponding instantaneous defect eigenstate throughout the process. Note, however, that the \emph{additional} zero-energy defect state arising from the sublattice imbalance of the square-root chain (see decoupled red site in the middle of Fig.~ \ref{fig:consecutive}) evolves independently and is not tied to the time evolution of the parent's defect states.

Choosing a quasi-adiabatic evolution (large total time $T$) of $H^{(1)}$ with initial (i) and final (f) angle configurations
\begin{equation}
\theta_{1,i} = \theta_{2,f} = \frac{\pi}{2}, \qquad
\theta_{2,i} = 0, \qquad
\theta_{3,i} = \lambda, \qquad
\theta_{3,f} = \gamma, \qquad
\theta_{1,f} = \frac{\pi}{2} - \lambda,
\label{eq:angle_schedule_1}
\end{equation}
where $\gamma\in\{0,\pi\}$, the evolution of the defect states is shown in Fig.~\ref{fig:transfer_states} (a more detailed account of the process is provided in Appendix \ref{app:transfer}).
\begin{figure}[ht!]
 \centering  
 \includegraphics[
width=0.9\textwidth]{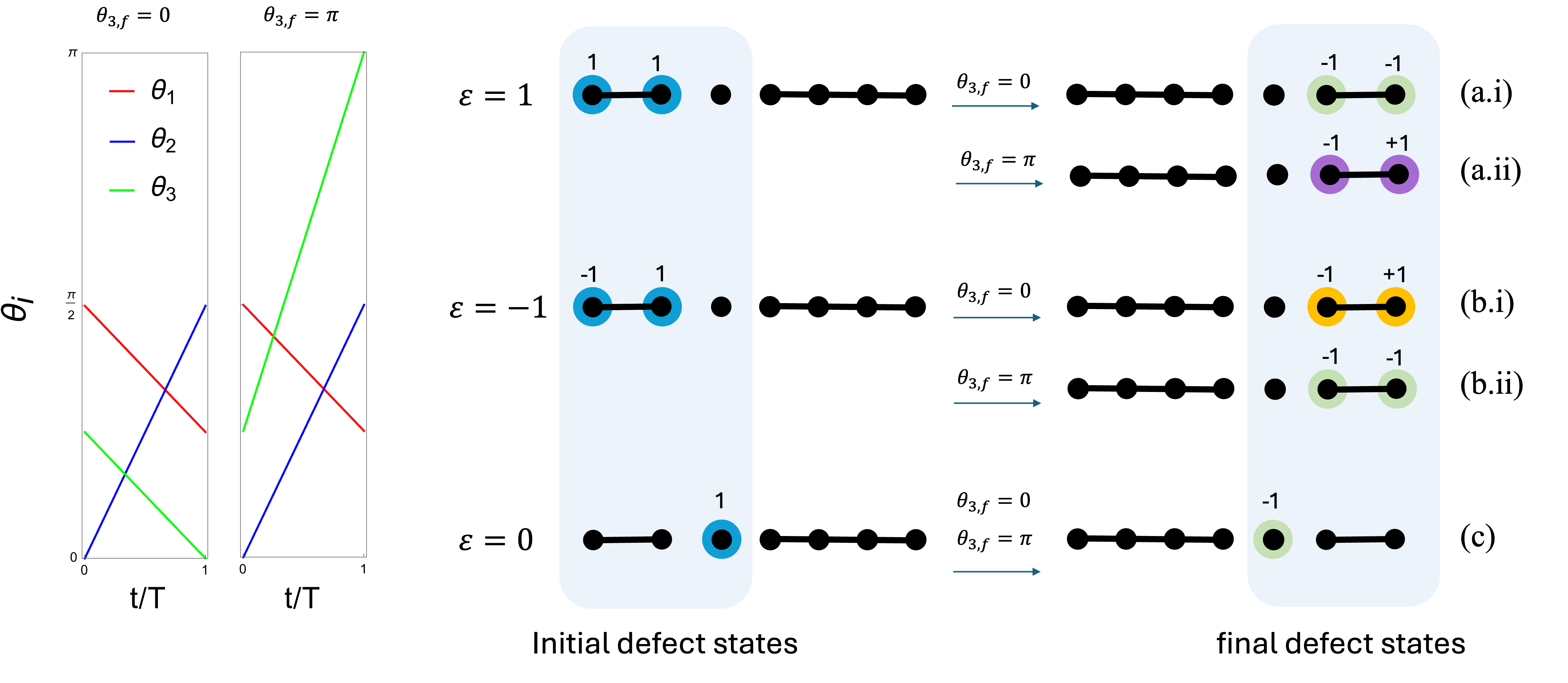}
  \caption{Simultaneous quasi-adiabatic quantum transfer of three energy states ($\varepsilon = 1$, $-1$, and $0$) in a Matryoshka chain. Panels (a–c) correspond to $\varepsilon = 1$, $-1$, and $0$, respectively, with each panel showing two scenarios: (i) $\theta_{3,f} = 0$ and (ii) $\theta_{3,f} = \pi$. The normalization constant was omitted for clarity. The left subplot illustrates the time evolution of angles $\theta_1$, $\theta_2$, and $\theta_3$ (red, blue, green lines), while the middle and right subplots depict the initial and final defect states.}
  \label{fig:transfer_states}
\end{figure}
For the defect states of energy $\varepsilon =\pm 1$, the relative signs within one sublattice follow those of the parent chain, while the relative signs in the other sublattice are fixed by the choice $\gamma=0$ or $\pi$. The zero-energy defect state has support only on the residual sublattice and is therefore unrelated to the SSH defect state.
The protocol thus transfers the Matryoshka defect states identically, the only additional effect being a global phase factor that depends on the energy of the transferred state. The three defect states $\varepsilon=\{+1,-1,0\}$ can therefore be moved simultaneously from one end of the chain to the other without loss of fidelity, provided the evolution is slow enough. By linearity, any superposition of the three states (i.e., a three-component qudit state) is transferred as well (with additional relative dynamical phases between components reflecting the energy difference of the edge states in different energy gaps). As we show in Sec.~\ref{sec:gates}, this simultaneous transfer is what enables generalized qudit quantum gates.

Note that, as illustrated by the highlighted region in Fig.~\ref{fig:consecutive}, the defect of $H^{(P)}$, with $P>1$, matches the full chain of $H^{(P-1)}$. Moreover, the diconnected clusters of the defect are interchangeable under adiabatic evolution.

\section{Quantum gates by Y-junction braiding}\label{sec:gates}

Gate-operation scalability is fundamental as the number of qubits grows in a quantum computer implementation, with decoherence times that exceed the time required to perform operations on large qubit numbers \cite{divincenzogates,Estarellas2018,Nielsen2010}. This requirement poses a challenge, since the quantum gate must act slowly to ensure quantum transfer is performed under adiabatic conditions, as discussed in the previous section. 
We address this by increasing the \emph{operation density} (the number of logical operations completed per physical device per unit time) rather than by speeding up individual gates.

In this section we extend the Y-junction braiding approach of Boross \emph{et al.} \cite{Boross2019} to the Matryoshka chain, using defects as carriers of qudit states in a Y-junction architecture \cite{Ezawa2020a}, and we test the resulting gates under three types of disorder.

In topological quantum computing, braiding is the adiabatic exchange of defects, implementing (generally non-commuting) unitary operations \cite{Dennis2002}. Because two defects cannot be exchanged along a single one-dimensional chain, a Y-junction architecture is used \cite{Ezawa2020a}. The hopping amplitudes within each Y-junction leg are tuned adiabatically so that the defects are driven around the junction without changing their energy. The Y-junction geometry lets one defect remain stationary while the other traverses the junction, completing the exchange.

\begin{figure}[ht!]
 \centering  
 \includegraphics[
width=0.9\textwidth]{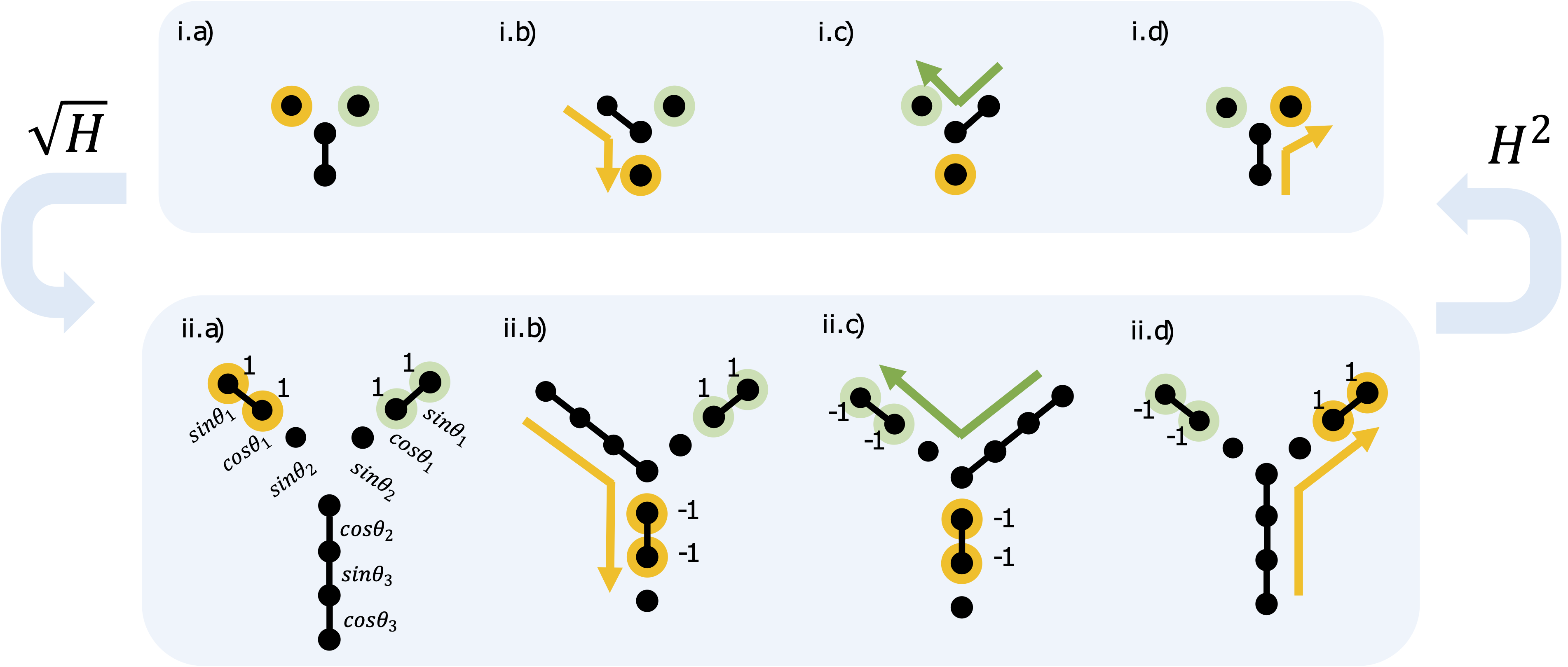}
  \caption{Braiding process in a (i) SSH-like and (ii) sine-cosine Y-junction. (i.a) Initial state with yellow (defect 1) and green (defect 2) circles. (i.b) and (i.c) show adiabatic transfer along semi-chains, with arrows indicating direction and the phase factors accumulated during transfer. (i.d) Final configuration after defect exchange. (ii.a)--(ii.d) depict the corresponding square-root Matryoshka chain parameters and phase accumulations. The gate operation corresponds to the transfer operator of Eq.~\eqref{eq:OP}, evaluated for the Matryoshka sine--cosine chain with $\theta_{3,f} = 0$.}
  \label{fig:braiding}
\end{figure}

Figure~\ref{fig:braiding}  (top row) illustrates the braiding protocol. The defect states undergoing braiding are drawn in distinct colors: the left defect state $\ket{L}$ in yellow and the right defect state $\ket{R}$ in green. These protected defect states span the computational space, that is,  the qubit basis is $\{\ket{R}, \ket{L}\}$. During each transfer step the defect states energy remains at $\varepsilon =0$ due to the chiral symmetry while acquiring path-dependent phase factors. This phase accumulation is what makes the braiding implement non-trivial unitary operations. In the parent SSH chain the braid maps $\ket{L}\to\ket{R}$ and $\ket{R}\to-\ket{L}$, i.e., a $Y$-type (bit-and-phase flip) operation.

\begin{figure}[ht!]
	\centering  
	\includegraphics[width=0.6\textwidth]{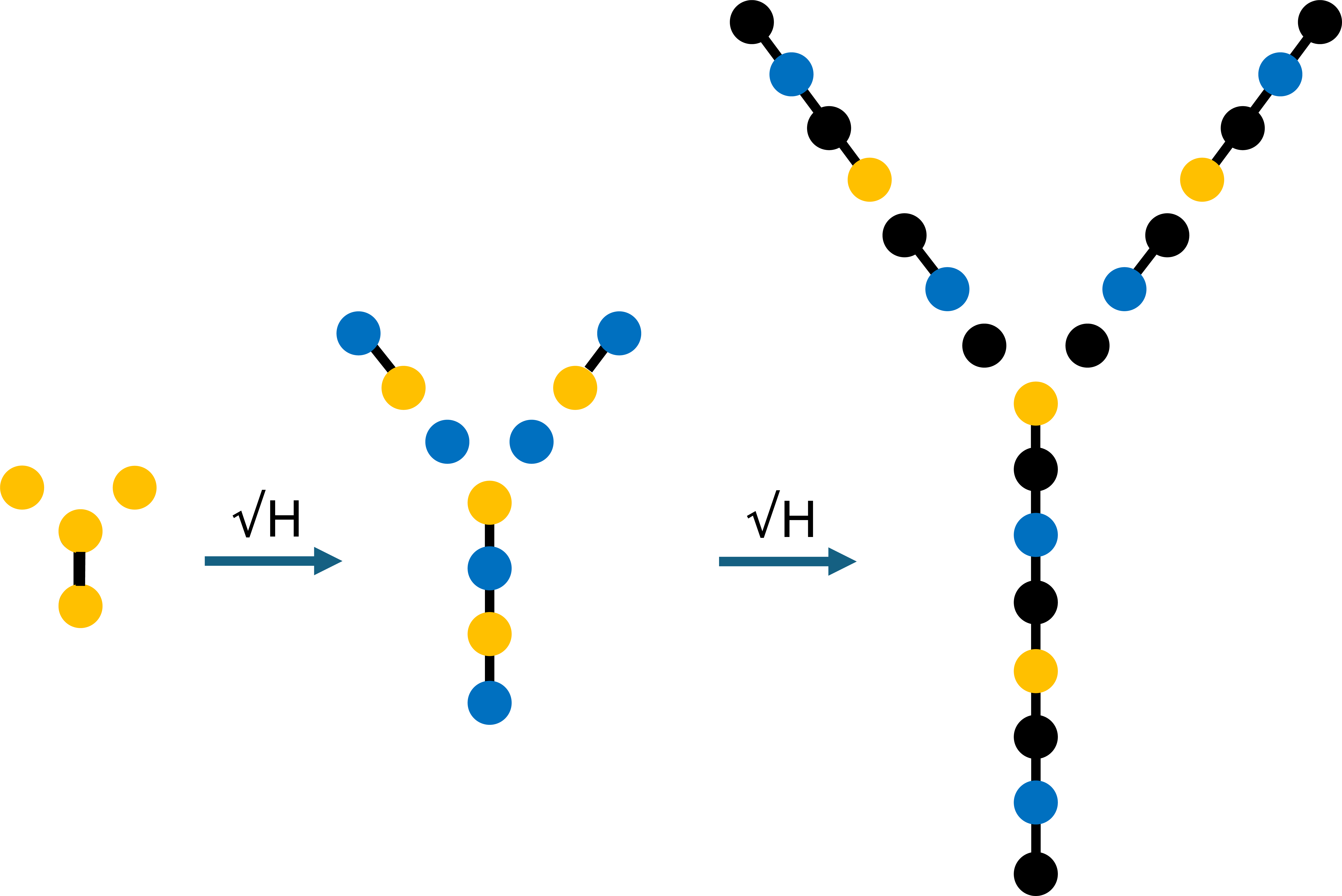}
	\caption{Scalable Y-junction quantum gate design using Matryoshka-type chains. Successive square-root operations ($\sqrt{H}$) generate higher-order structures with generalized defects, each supporting a number of distinct states that grows with the chain order $N$ ($N = 1, 2, 3, \dots$). The diagram illustrates the progressive nesting of defects across three hierarchical levels.}
	\label{fig:gate_design}
\end{figure}

Following Sec.~\ref{sec:transfer}, we replace the standard SSH Y-junction with a sine--cosine Matryoshka Y-junction, the square root of the former (Fig.~\ref{fig:braiding}, bottom row). Each defect now supports three eigenstates, with energies $\varepsilon = \pm 1$ and $\varepsilon = 0$.
Figure \ref{fig:braiding} illustrates the braiding of the $\varepsilon = 1$ defect states for $\theta_{3,f} = 0$ (a $\pi$ phase factor is accumulated after each two-leg state transfer); see Appendix A for a detailed description of the  hopping angles evolution in each transfer. The phases acquired in this braiding are inherited directly from the phases accumulated by the parent system in its own braiding process.

The hierarchical nesting of generalized defects across successive square-root orders is illustrated in Fig.~\ref{fig:gate_design}.
For the $P=1$ case with three-site defects, the quantum information is encoded in the six-dimensional defect subspace (a qudit of dimension $d=6$) spanned by 
\newline
\begin{minipage}{0.5\textwidth}
	\small
\begin{align}
\ket{1} &= \ket{\varepsilon=1; L} = \frac{1}{\sqrt{2}} \begin{bmatrix} 1 & 1 & 0 & 0 & 0 & 0 \end{bmatrix}^{T} \\
\ket{2} &= \ket{\varepsilon=1; R} = \frac{1}{\sqrt{2}} \begin{bmatrix} 0 & 0 & 0 & 0 & 1 & 1 \end{bmatrix}^{T} \\
\ket{3} &= \ket{\varepsilon=-1; L} = \frac{1}{\sqrt{2}} \begin{bmatrix} 1 & -1 & 0 & 0 & 0 & 0 \end{bmatrix}^{T} 
\end{align}
\end{minipage}
\hfill
\begin{minipage}{0.5\textwidth}
	\small
\begin{align}
\ket{4} &= \ket{\varepsilon=-1; R} = \frac{1}{\sqrt{2}} \begin{bmatrix} 0 & 0 & 0 & 0 & 1 & -1 \end{bmatrix}^{T} \\
\ket{5} &= \ket{\varepsilon=0; L} = \begin{bmatrix} 0 & 0 & 1 & 0 & 0 & 0 \end{bmatrix}^{T} \\
\ket{6} &= \ket{\varepsilon=0; R} = \begin{bmatrix} 0 & 0 & 0 & 1 & 0 & 0 \end{bmatrix}^{T}
\end{align}
\end{minipage}
\newline
\newline
\noindent where the components of each 6-dimensional vector give the amplitudes at the highlighted defect sites of   Fig.\ref{fig:transfer_states}, with entries 1–3 referring to the three sites of the left defect and entries 4–6 to the three sites of the right defect, following a left to right ordering in both.

For the Matryoshka braiding of Fig.~\ref{fig:braiding}, the transfer operator expressed in this defect basis is block-diagonal in the energy sectors,
\begin{equation}
\mathbf{OP} = \begin{bmatrix}
e^{-it/\hbar} \begin{pmatrix} 0 & -1 \\ 1 & 0 \end{pmatrix} & \boldsymbol{0_{2\times2}} & \boldsymbol{0_{2\times2}} \\
\boldsymbol{0_{2\times2}} & e^{it/\hbar} \begin{pmatrix} 0 & 1 \\ 1 & 0 \end{pmatrix} & \boldsymbol{0_{2\times2}} \\
\boldsymbol{0_{2\times2}} & \boldsymbol{0_{2\times2}} & \begin{pmatrix} 0 & -1 \\ 1 & 0 \end{pmatrix}
\end{bmatrix} =
\begin{bmatrix}
e^{-it/\hbar} \, Y_{\varepsilon=1} & \boldsymbol{0_{2\times2}} & \boldsymbol{0_{2\times2}}\\
\boldsymbol{0_{2\times2}} & e^{it/\hbar} \, X_{\varepsilon=-1} & \boldsymbol{0_{2\times2}}\\
\boldsymbol{0_{2\times2}} & \boldsymbol{0_{2\times2}} & Y_{\varepsilon=0}
\end{bmatrix},
\label{eq:OP}
\end{equation}
where $t$ is the braiding time and $Y_{\varepsilon=1}$, $X_{\varepsilon=-1}$, and $Y_{\varepsilon=0}$ act on the $\varepsilon=+1$, $-1$, and $0$ sectors, respectively. 

The block-diagonal structure is the key advantage of the Matryoshka model over a SSH chain: a \emph{single} braiding event can apply one gate operation in \emph{each} energy sector simultaneously. In the $P=1$ case, the operation density per physical device is thus three times that of the SSH chain, and for general order $P$ the number of independent defect sectors that can be addressed at once is $2^{P+1}-1$ [one per protected gap, see Eq.~\eqref{eq:n_edge}].

\subsection{Robustness of the gate to disorder}\label{subsec:disorder}

\begin{figure}[ht!]
	\centering  
	\includegraphics[width=1\textwidth]{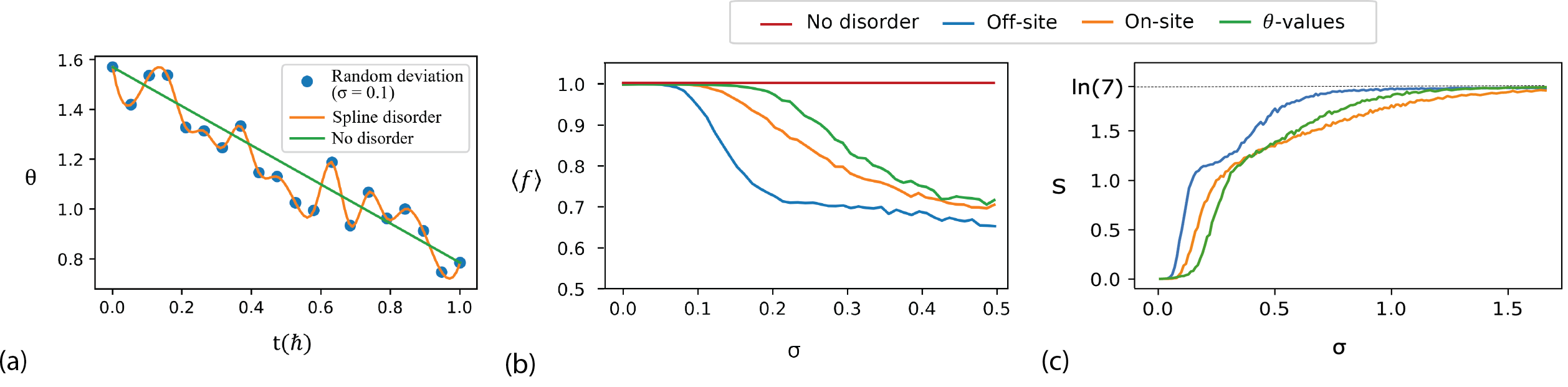}
	\caption{(a) Example of smooth angle disorder introduced by spline interpolation over 20 random deviations (blue dots) sampled from a normal distribution with $\sigma = 0.1$. The green curve is the  disorder-free evolution while the orange curve corresponds to the interpolated noisy evolution. (b) Results for the transfer amplitude $f=|\bra{\Psi_f}\ket{\Psi_t}|$  and (c)  for the entropy, averaged over  $N = 1000$ independent realizations of the transfer of $\ket{\Psi_0} = (\ket{1} + \ket{2})/\sqrt{2}$ to $\ket{\Psi_f} = -(\ket{6} + \ket{7})/\sqrt{2}$, using the hopping configuration with $\gamma=0$ from Fig.~\ref{fig:transfer_states} and smoothed (spline-interpolated) disorder, for a total transfer time $T=1$ in units of $\hbar$.}
	\label{fig:disorder}
\end{figure}

An evolution is adiabatic when every Hamiltonian parameter varies on a timescale that is long when compared with that set by the inverse of the smallest instantaneous level spacing, so that transitions between instantaneous eigenstates remain negligible \cite{Born1928}. For the transfer to have high fidelity, the (i) on-site (diagonal), (ii) off-diagonal (hopping), and (iii) angle parameters $\theta_j$ must therefore all change smoothly in time. Since perturbations acting on a physical device are expected to be slowly varying, smooth disorder is also the realistic one. Figure~\ref{fig:disorder}(a) shows such disorder generated by spline-interpolating 20 random deviations sampled from a normal distribution with standard deviation $\sigma = 0.1$ around the original $\theta(t)$ curve (green), together with two noisy evolutions (orange, blue).

Because the gate consists of three consecutive transfers applied one after another, each stage accumulates errors of the same character, so that fidelity decays and entropy grows faster in the full gate than in any single transfer. The disorder effects can hence be read off from a single transfer; we illustrate them using the top quantum transfer of Fig.~\ref{fig:transfer_states}(a.i) (corresponding to $\theta_{3,f} = 0$), which has a three-component defect in a seven-site chain. Labelling the sites from one to seven starting at the left end, the initial state is
$\ket{\Psi_0} = (\ket{1} + \ket{2})/\sqrt{2}$,
which evolves into $\ket{\Psi_f} = -(\ket{6} + \ket{7})/\sqrt{2}$ after the transfer operation.

Fidelity and entropy are evaluated numerically on the spline-smoothed perturbations and averaged over $N = 1000$ disorder realizations at each value of $\sigma$ [Fig.~\ref{fig:disorder}(b,c)]. The fidelity $F$, measuring the overlap between the expected state $\ket{\Psi_f}$ and the actually transferred state $\ket{\Psi_t}$, is written as $F = f^2= |\bra{\Psi_f}\ket{\Psi_t}|^2$.

Of the three disorder types, hopping disorder (ii) produces the fastest fidelity decay, while the correlated disorder of the angles $\theta_j$ (iii) is the most stable and keeps the fidelity high over the widest range of disorder strength.This is expected behavior since correlated disorder preserves the chiral symmetry in the parent lattice (see Appendix B, in particular Fig.~ \ref{fig:disorder_increase}). The von Neumann entropy [$S(\rho) = -\text{Tr}(\rho \ln \rho)$ where $\rho_t= \ket{\Psi_t}\bra{\Psi_t}$] behaves complementarily: it grows rapidly toward its maximum value, faster for case (ii) than for case (iii). The maximum is $\ln 7$, corresponding to the maximally mixed state with equal weight on all seven basis states of the chain.

\section{Quantum memories}\label{sec:memory}

In this section we construct a quantum memory based on the Matryoshka model by extending the protocol of \cite{palyi2018ssh}. Because the Matryoshka construction adds protected energy gaps and hence protected localized eigenstates, a single chain can store a larger set of qubits (or equivalently a high-dimensional qudit) than an SSH chain of the same length. The set of chiral symmetries at successive orders $0, 1, \dots, P-1$  plays a crucial role in preserving these topological properties throughout the whole chain \cite{Dias2021}.

\begin{figure}[ht!]
 \centering  
 \includegraphics[
width=0.95\textwidth]{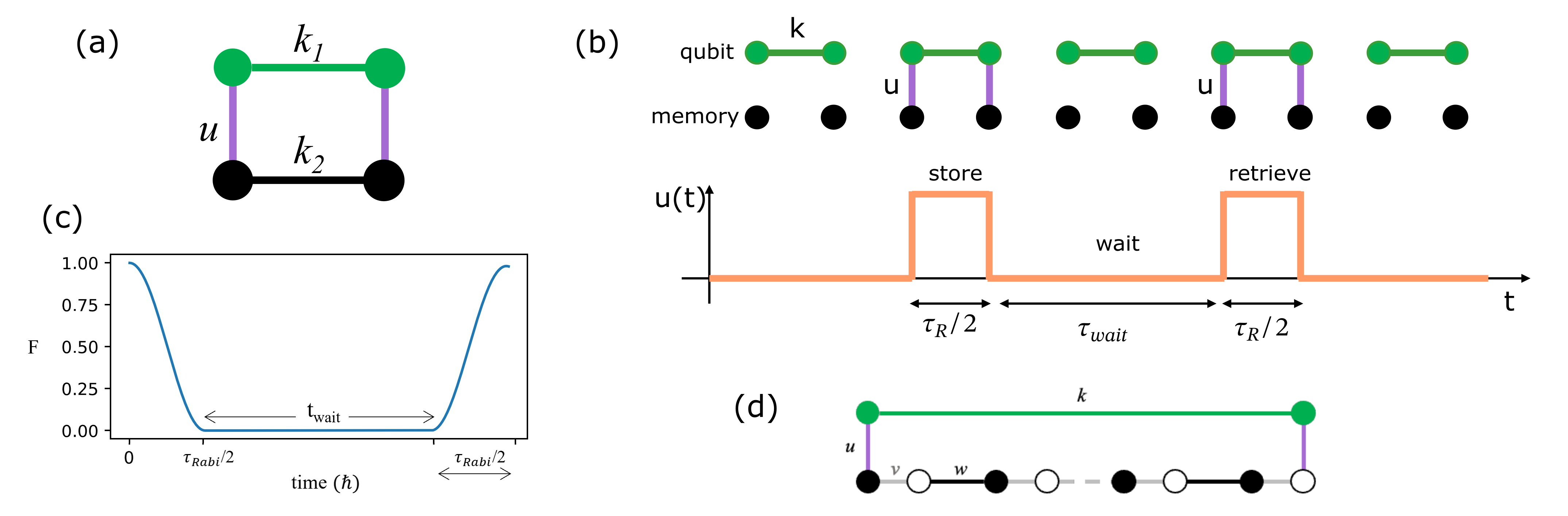}
  \caption{(a) Dot memory model showing two qubits with internal hopping terms $k_1$ (green line) and $k_2$ (black line). The qubits are connected through a coupling $u$ (purple line). (b) Schematic representation from \cite{palyi2018ssh} of the qubit‑memory transfer process. The particle is first stored in the memory (black sites) after a Rabi half‑period $\tau_R/2$, remains there during a waiting time $t_{wait}$, and is then retrieved back into the qubit (green sites) over another interval of $\tau_R$. (c) Expected fidelity evolution during a perfect quantum state transfer of a qubit into a memory. The qubit state transfers into the memory after a period $t_{R}/2$ and stays secured in the memory for a period $t_{wait}$. Then, it is transferred back after $t_{R}/2$. (d) Memory transfer model using one qubit connected to an SSH chain with an even number of sites. The qubit has internal hopping $k$ (green line) and it is connected to the memory with hopping $u$ (purple lines).}
  \label{fig:memory_model}
\end{figure}

We follow the transfer protocol of \cite{palyi2018ssh} to move a qubit into and out of the memory. The qubit is represented by two sites with an internal transition amplitude $k$. Its computational basis is $\ket{0}=\ket{10}$ and $\ket{1}=\ket{01}$, i.e., the particle localized on the left or right site. The qubit is initially disconnected from the memory chain (a two-site chain in the simple case of  Fig.~\ref{fig:memory_model}(a)) and contains a single particle. The chain couples to the qubit via a hopping $u$. The qubit (green sites) and the memory (black sites) may have their own internal small hopping term, $k_1$ and $k_2$, respectively, that lead to coherence loss.  At the start of the protocol the coupling is off and the particle resides in the qubit.
When the coupling is turned on, the particle undergoes Rabi oscillations between the qubit and the memory with period $\tau_{R}=\hbar/u$. After a half-period $t=\tau_{R}/2$, the particle has migrated into the memory, the coupling is turned off and therefore the particle remains there for a waiting time $t_{\text{wait}}$. To retrieve the qubit, the coupling is reactivated and a second half-period of Rabi dynamics returns the particle to the qubit. The sequence is illustrated in Fig.~\ref{fig:memory_model}(b) in the case of an ideal memory.

Because the entire protocol is time-independent during each Rabi half-period, a general qubit state $\ket{\psi}=\alpha\ket{0}+\beta\ket{1}$ is preserved exactly, that is, the amplitudes $\alpha$ and $\beta$ are transfered to the two memory sites  with identical evolution, so their relative phase (the coherence) is unchanged. 
For the clean system,   the fidelity starts at $F=1$, drops to $F=0$ during the first half-period (the state is fully inside the memory), and returns to $F=1$ after the retrieval half-period, with the output state identical to the input [see Fig.~\ref{fig:memory_model}(c)]. 

In \cite{palyi2018ssh} an SSH chain was proposed as the memory, with the qubit connected to the end sites of the chain via hopping $u$ [see Fig.~\ref{fig:memory_model}(d)] Simulations showed that the fidelity is maximized in the dimerized limit ($v=0$), where the edge states are localized and protected ($F=1$). Away from this limit ($v>0$) the edge states spread onto neighboring sites, reducing transfer efficiency during $\tau_R/2$ and causing leakage into dispersive states. This produces short-time, small-amplitude oscillations (dispersive transfer) and long-time, large-amplitude oscillations (due to the hybridization of the left and right edge states), with the fidelity eventually reaching zero. Stronger transfer coupling (larger $u$) improves efficiency and leads to a  reduced fidelity decay for longer storage times.

\begin{figure}[ht!]
 \centering  
 \includegraphics[
width=0.8\textwidth]{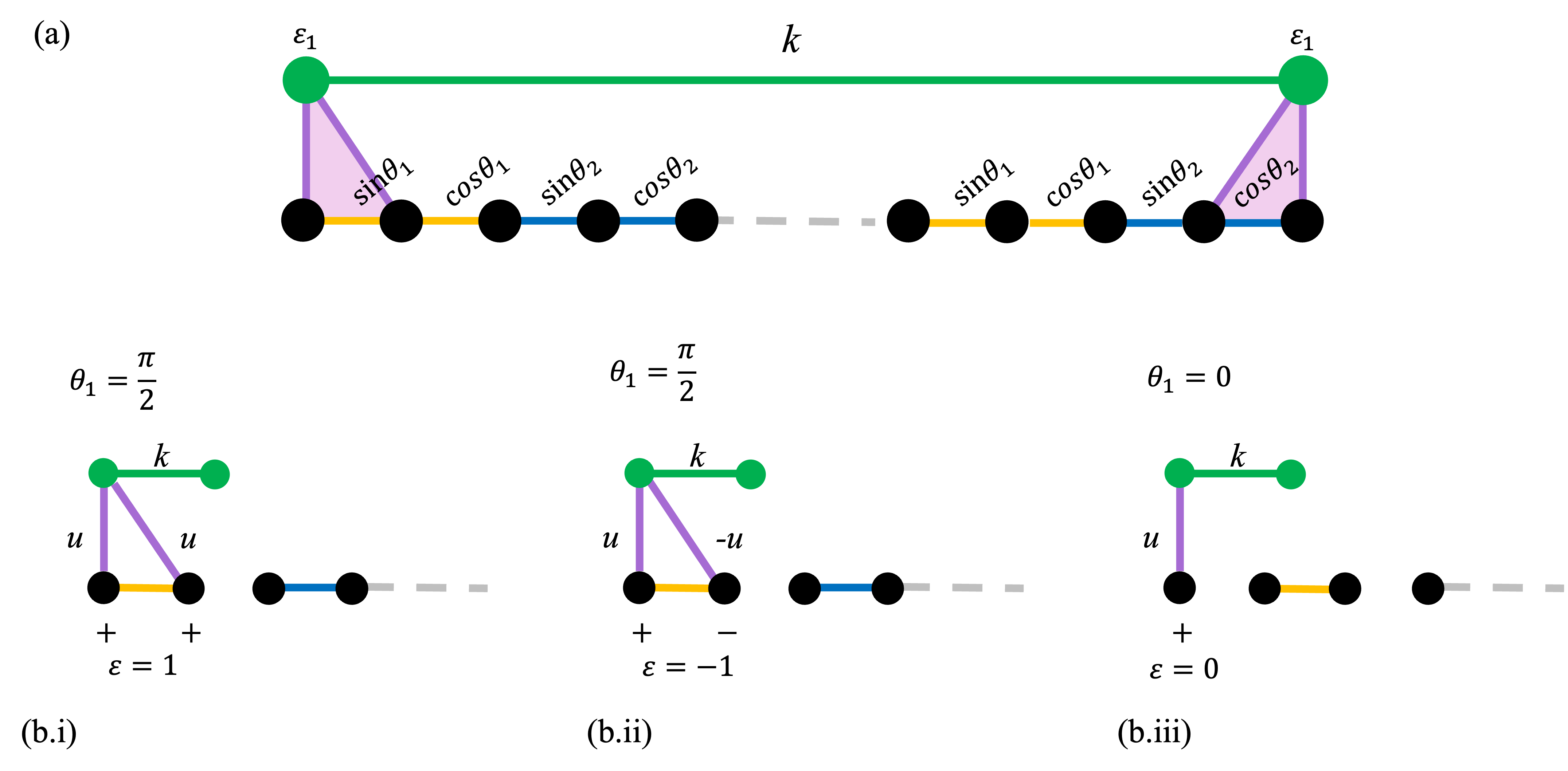}
  \caption{(a) Memory transfer model using one qubit connected to a Matryoshka chain of type $P=1$. The qubit has internal hopping $k$ (green line) and its connection to the memory depends on the energy of the edge state that will be used as memory. (b) Transferring method to different energy states. (b.i) Transfer of the qubit to the eigenstate with energy $\varepsilon=1$ (symmetric combination). Requires $\theta_1=\pi/2$. (b.ii) Transfer of the qubit to the eigenstate with energy $\varepsilon=-1$ (anti‑symmetric combination). Requires $\theta_1=\pi/2$. (b.iii) Transfer of the qubit to the eigenstate with energy $\varepsilon=0$. Requires $\theta_1=0$.}
  \label{fig:matryoska_memory}
\end{figure}

\begin{figure}[ht!]
 \centering  
 \includegraphics[
width=0.9\textwidth]{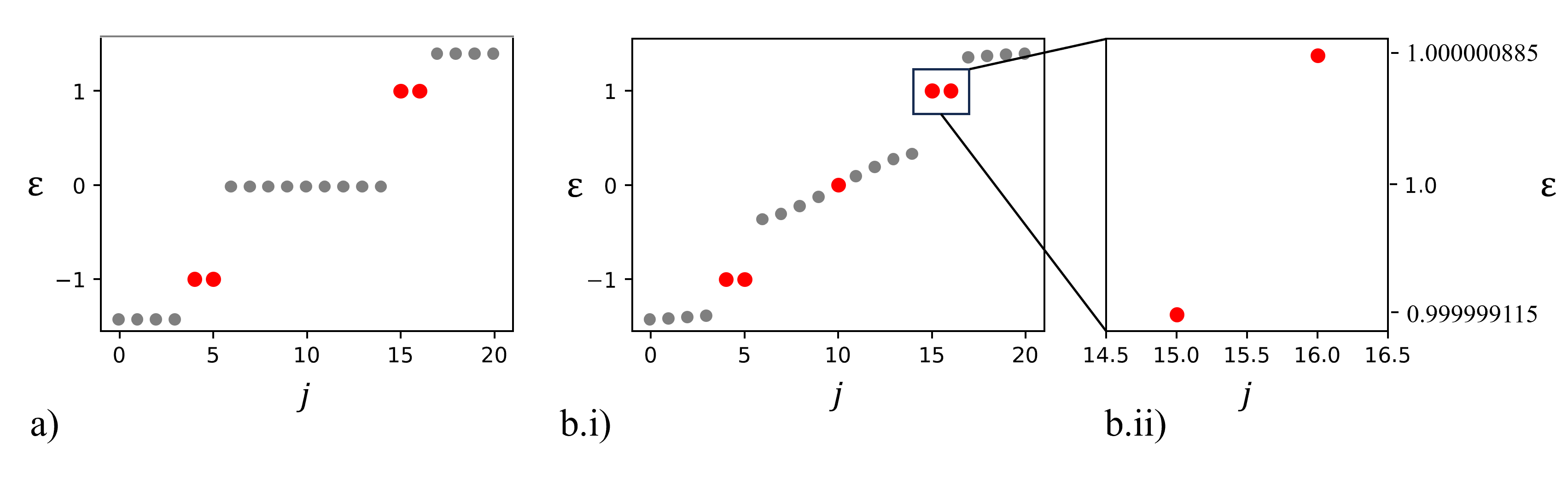}
  \caption{Energy spectrum of a Matryoshka chain with 21 sites ($P=1$). (a) Edge-dimerized configuration with $\theta_1=\pi/2$ and $\theta_2=0$, showing degenerate edge states at energy $\varepsilon=\pm1$. (b.i) Non-edge-dimerized configuration with $\theta_1=5\pi/12$ and $\theta_2=\pi/2-\theta_1$, where the degeneracy is lifted and state energies shift. (b.ii) Zoomed view of states near $\varepsilon=\pm1$, revealing a small energy splitting $\Delta\varepsilon \approx 1.77\times10^{-6}$ between formerly degenerate levels.}
  \label{fig:matryoska_spectrum}
\end{figure}

Following Secs.~\ref{sec:model}--\ref{sec:gates}, we replace the SSH memory with a Matryoshka-type chain and keep one qubit with internal hopping $k$ connected to the tips of the chain with hopping $u$. We use the simplest non-trivial structure, of order $P=1$, as illustrated in Fig.~\ref{fig:matryoska_memory}. The qubit-to-chain hopping parameters are chosen according to the edge state in which the qubit is to be stored [see Fig.~\ref{fig:matryoska_memory}(b.i--iii)].
A $P=1$ chain with an odd number of sites, in the edge-dimerized limit ($\theta_1=\pi/2$, $\theta_2=0$, $\cos\theta_3=0$), has four non-zero-energy edge states: two with energy $\varepsilon = -1$ (anti-symmetric states localized at the first and last dimers) and two with energy $\varepsilon = 1$ (symmetric states localized at the first and last dimers) [see Fig.~\ref{fig:matryoska_spectrum}(a)]. 

Two of the eigenstates with energy $\varepsilon= \pm 1$ will be localized at the leftmost part of the chain. To transfer an initial state $\ket{\Psi}=\ket{0}$ of the qubit (corresponding to the particle localized at the left site), two connections should be present, so it transfers with equal weights either to the symmetric or to the anti‑symmetric states localized at the leftmost dimer [see Fig.~\ref{fig:matryoska_memory}(b.i) and (b.ii)]. For the $\theta_1 = 0$ case, there is one $\varepsilon =0$ eigenstate fully localized at the first site of the chain, so the transfer only requires one connection [see Fig.~\ref{fig:matryoska_memory}(b.iii)].

The cases just described assume dimers at the chain edges.  In  Fig.~\ref{fig:matryoska_spectrum}(a), we show the energy spectrum of the dimerized $P=1$ open chain with $\theta_1=\pi/2$, $\theta_2=0$, and $N=21$ sites. Each edge state (red dot) has an exact double degeneracy. This degeneracy is present due to the extra site from a incomplete unit cell at the right edge of the chain,   shown in   Fig.~\ref{fig:matryoska_spectrum}(a), that guarantees that, when squaring the $P=1$ Matryoshka chain, one obtains a parent Hamiltonian with inversion symmetry.

\begin{figure}[ht!]
 \centering  
 \includegraphics[
width=0.9\textwidth]{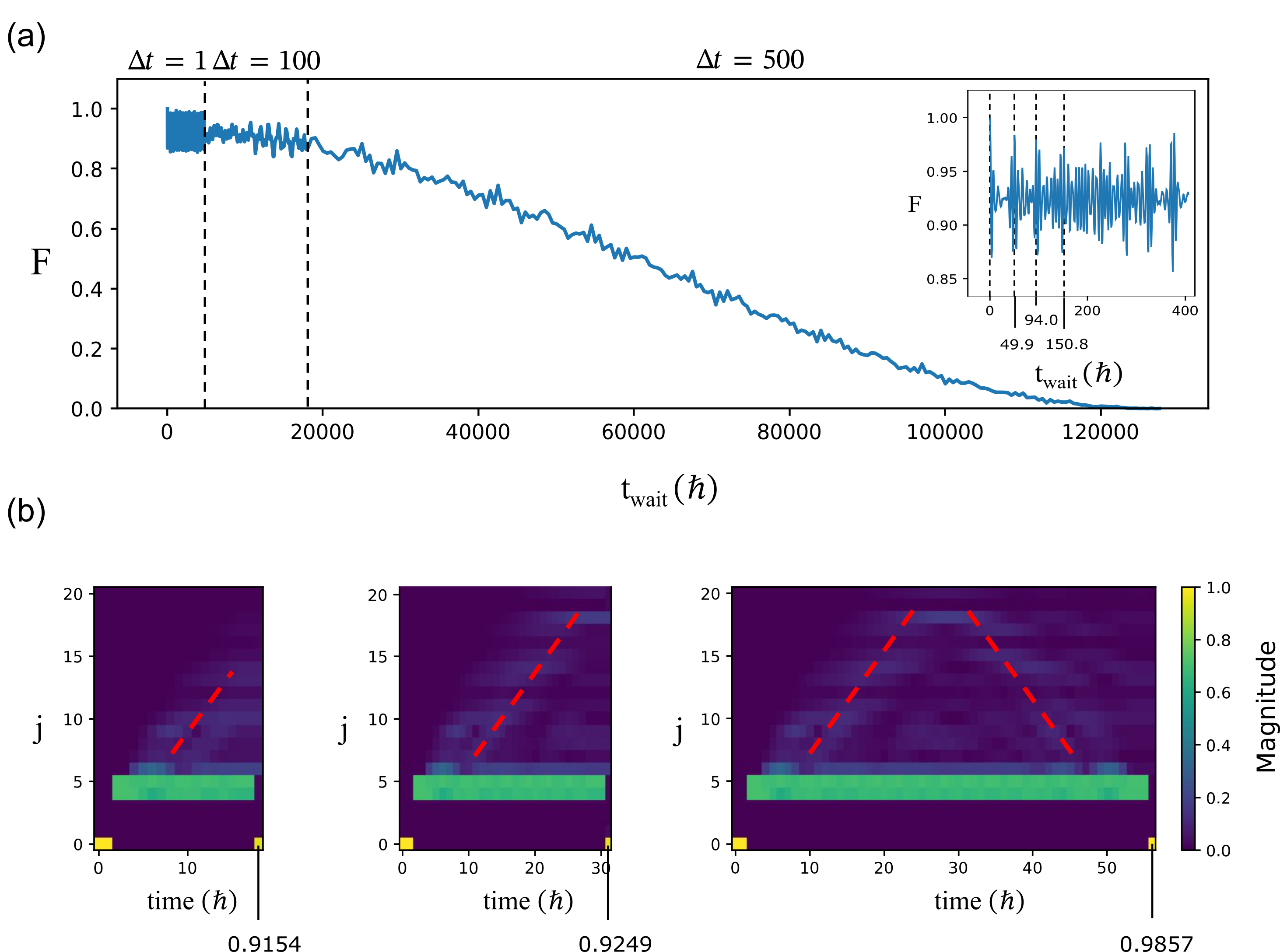}
  \caption{(a) Fidelity $F$ versus storage time $t_{\text{wait}}$ in a Matryoshka chain with 21 sites ($P=1$, $\theta_1 = \pi/12$, $\theta_2 = \pi/2 - \theta_1$). The plot displays half of the large-amplitude fidelity oscillation, corresponding to quantum state transfer from the left edge state to the right edge state. Three regions with different sampling times are separated by vertical dashed lines. Inset: zoomed view near $t_{\text{wait}} = 0$, highlighting the first four peaks (including $t_{\text{wait}} = 0$) where maximum fidelity is reached, indicated by vertical dashed lines. (b) Density plots of the quantum state for storage times before and close to the second fidelity peak. The dashed red line traces the path of the state probability amplitudes as it shuttles between the left and right edge states over the storage interval. The initial state $|\Psi_0\rangle = |k=0\rangle$ is symmetrically transferred to the leftmost two sites, stored for duration $t_{\text{wait}}$, then transferred back to the qubit with expected result $|\Psi_f\rangle = |k=0\rangle$. Final fidelity values are shown in the bottom-right corners of each image.}
  \label{fig:fidelity_oscillation}
\end{figure}

For other hopping parameters the edge states are no longer fully localized at the leftmost and rightmost pairs of sites and instead carry exponentially decaying weight into the bulk.
Stepping slightly away from the edge-dimer limit, e.g.\ to $\theta_1=5\pi/12$ and $\theta_2=\pi/2-\theta_1$, the degenerate edge states acquire a small energy splitting [see Fig.~\ref{fig:matryoska_spectrum}(b)], reflecting hybridization between the left and right edge states. The stored state then undergoes Rabi oscillations with period $\tau_R=\pi/\Delta\varepsilon$, so the fidelity, which would stay at $F=1$ for any $t_{\text{wait}}$ in the dimerized limit, now oscillates with $t_{\text{wait}}$. This mirrors the behavior found for the SSH chain away from its dimerized limit \cite{palyi2018ssh}. Because of the edge-state spreading, part of the amplitude leaks into dispersive states during the transfer of Fig.~\ref{fig:matryoska_memory}(a.i)--(a.ii), and the final fidelity varies with the storage time $t_{\text{wait}}$, as shown in Fig.~\ref{fig:fidelity_oscillation}(a), showing short-time small-amplitude and long-time large-amplitude oscillations.
Higher fidelity is recovered if the state is kept in memory for a duration in which an integer number of oscillations completes, so that most of the amplitude returns to the initial two sites. The hybridization-induced splitting between the symmetric and anti-symmetric states [see Fig.~\ref{fig:matryoska_spectrum}(b.ii)] directly predicts the long-time fidelity decay, with a half-period $\tau_R/2 = \pi/(2\Delta\varepsilon)$.
The inset of Fig.~\ref{fig:fidelity_oscillation}(a) shows that the fidelity reaches periodic maxima, the first at $t_{\text{wait}} = 49.9$. This period matches the time evolution of the state-density localization in Fig.~\ref{fig:fidelity_oscillation}(b). The dashed red line traces the state amplitudes shuttling between the left and right edge states, and it returns to the initial (left-edge) configuration after an interval of about 50 time units, consistent with the first fidelity peak.

\section{Conclusion}\label{sec:conclusion}

Scalability remains one of the central challenges in building practical quantum computers, due to the physical resource overhead and the cost of error mitigation. This work proposes a route toward scalable quantum information processing based on a hierarchical lattice structure, the Matryoshka-type sine--cosine chain, that encodes a high-dimensional qudit (or several qubits) within a \emph{single} physical device. Unlike architectures that allocate one physical system per qubit, the $2^n$-root construction lets multiple protected edge states and energy gaps coexist in one chain, reducing the number of independent devices per logical degree of freedom.

We showed that: (i) the adiabatic defect-transfer step underlying the poor-man's topological gate of Boross \emph{et al.} \cite{Boross2019} carries over unchanged to the Matryoshka chain and transfers all defect states simultaneously; (ii) a Y-junction braiding protocol then composes single-qubit gate operations in every energy sector at once, raising the operation density per device. These gates exhibit partial robustness against local perturbations due to the chiral symmetry present at each order of the hierarchical square-root sequence leading to the $P=n$ chain; (iii) the same edge states serve as a quantum memory whose capacity grows proportionally to the number of protected edge states (this number grows exponentially with the chain order $P$ at fixed physical length). 

We also integrated Matryoshka chains into quantum-memory architectures. We have shown that  memory fidelity remains high over extended storage periods, and the capacity scales exponentially with $P$, enabling multi-qubit or qudit storage within a single device.

Two caveats should be kept in mind. First, the protocols require \emph{dynamically tunable} hopping amplitudes (the adiabatic angle schedules of Sec.~\ref{sec:transfer}). A realistic implementation needs a platform with at least quasi-statically reconfigurable couplings. Second, the protection is partial. It is guaranteed by a sequence of chiral symmetries at the square-root steps and adiabaticity is required. Disorder strong enough to close a protected gap destroys this protection, as detailed in App.~\ref{app:robustness}. 

Experimental realization is feasible in photonic waveguides fabricated by femtosecond laser writing, which allow precise control over the hopping parameters and an effective simulation of the discrete quantum system \cite{Longhi2009,Perez2024,ViedmaPalomo2024,Zhou2020,Li2020}. Classical analogues such as topolectrical circuits \cite{Song2020,Liu2019,Lee2018} and acoustic lattices \cite{Yan2020,Xue2019,Zheng2019} offer additional platforms for testing the scalability of these architectures, particularly the multi-sector gate structure of Sec.~\ref{sec:gates}.

\section*{Acknowledgements} 

This work was developed within the scope of Portuguese Institute for Nanostructures, Nanomodelling and Nanofabrication (i3N) Projects No.~UIDB/50025/2020, No.~UIDP/50025/2020, and No.~LA/P/0037/2020, financed by national funds through the Funda\c{c}\~{a}o para a Ci\^{e}ncia e Tecnologia (FCT) and the Minist\'{e}rio da Educa\c{c}\~{a}o e Ci\^{e}ncia (MEC) of Portugal. 
A.M.M. acknowledges financial support from i3N through the work Contract No.~CDL-CTTRI-91-SGRH/2024.
G.F.M. acknowledges financial support from FCT through the grant BI/UI96/12758/2025.

\appendix
\section{Quantum transfer in sine--cosine chains}\label{app:transfer}

This appendix shows how quantum state transfer protocols based on three-site SSH chains also apply to Matryoshka-type chains, as their Hamiltonians preserve SSH eigenstates as sublattice amplitudes. We detail the adiabatic evolutions of the hopping parameters to move defects and acquire controlled phase shifts, demonstrating this is a scalable method for transferring a large number of quantum states within the topological framework.

The sine--cosine chains are bipartite; the Hamiltonian can therefore be written in the chiral basis $\{\{\ket{A_i}\},\{ \ket{B_i}\}\}$ where 
$\ket{A_i} = \ket{2(i-1)}$, $i=1,\cdots,N_A$, and  $\ket{B_i} = \ket{2i-1}$, $i=1,\cdots,N_B$,  with $N_A$ and $N_B$ being the number of sites of type $A$ and $B$, respectively. Therefore, in this basis one has 
\begin{equation}
	H= \begin{bmatrix} 0 & H_{AB} \\ H_{BA} & 0\end{bmatrix},
\end{equation}
The respective eigenvalue equation can also be written as $H_{AB} \ket{A}= \varepsilon \ket{B}$ and $H_{BA} \ket{B}= \varepsilon \ket{A}$, where $\ket{A} = \sum_{i=1}^{N_A} \alpha_i \ket{A_i}$ and  $\ket{B} = \sum_{i=1}^{N_B} \beta_i \ket{B_i}$ are vectors with non‑zero values only at the $A$ and $B$ sublattices, respectively. 

\begin{figure}[ht!]
 \centering  
 \includegraphics[
width=0.9\textwidth]{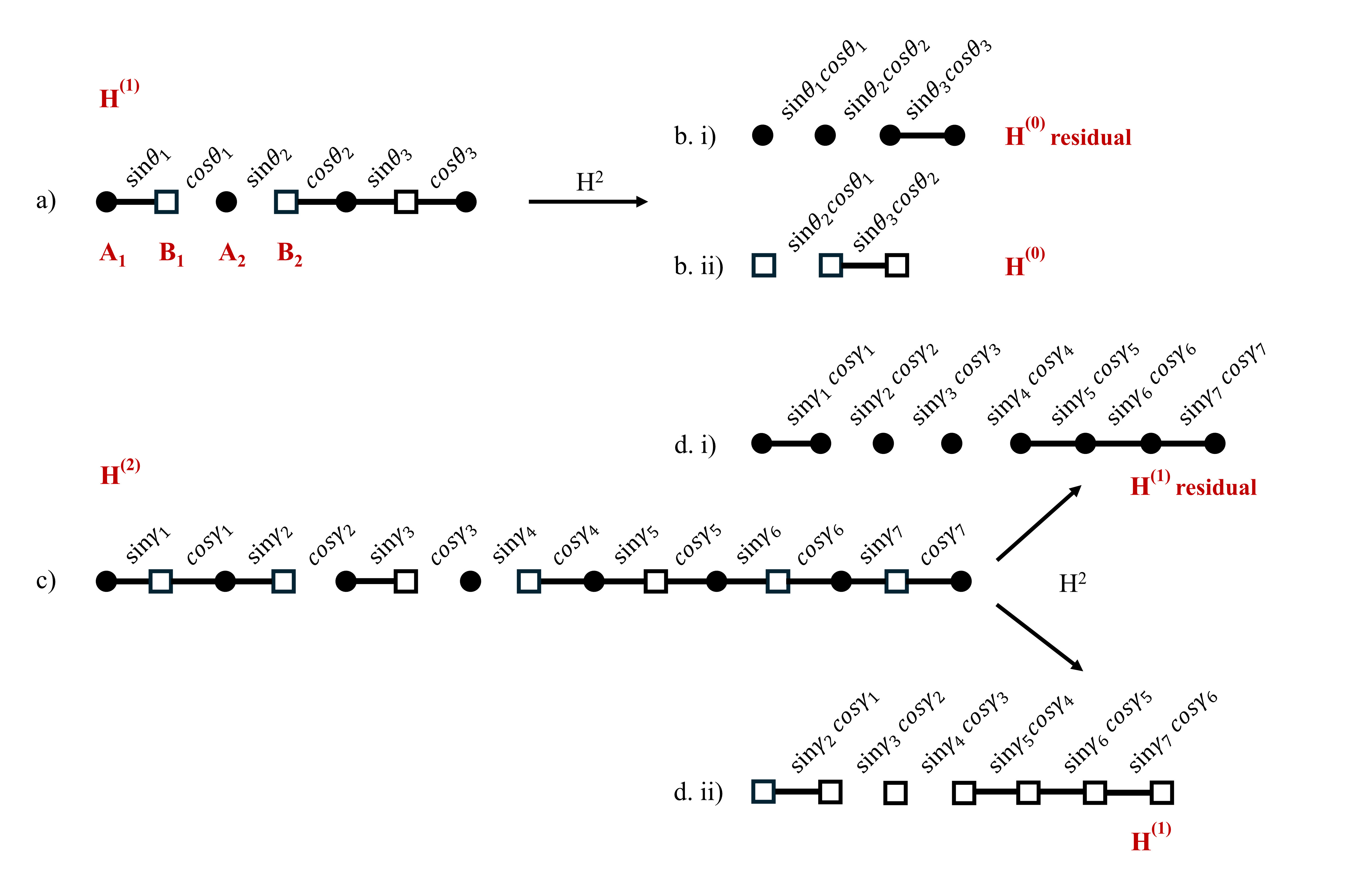}
  \caption{Recursive squaring of Matryoshka lattices. Squaring the Hamiltonian $H^{(1)}$ (a) yields a residual structure (b.i) and the parent structure (b.ii). Similarly, squaring $H^{(2)}$ (c) produces a parent structure (d.ii) identical to (a), and the residual structure (d.i). On-site terms are omitted.}
  \label{fig:recursive_squaring}
\end{figure}

Consider the square root of the three-site base structure of Fig.~\ref{fig:transfer_protocol}, denoted $H^{(1)}$; the defect is the group of three leftmost sites. Throughout the transfer, $H^{(1)}$ hosts the $\varepsilon=\pm1$ states that descend from the square root of the $H^{(0)}$ zero-energy state. Using the parent/square-root relation (illustrated in Fig.~\ref{fig:recursive_squaring}) and the form of the $H^{(0)}$ zero-energy state,
\begin{equation}
\ket{B} = \begin{bmatrix} \cos{\theta},& 0,& - \sin{\theta}\end{bmatrix}^T = \begin{bmatrix} \cos{\theta_2}\sin{\theta_3}, & 0,& -\cos{\theta_1}\sin{\theta_2} \end{bmatrix}^T,
\end{equation}
state $\ket{A}$ can be inferred from $\ket{B}$,
\begin{equation}
\ket{A} = \begin{bmatrix} \beta_1 \sin{\theta_1},& \beta_1 \cos{\theta_1} + \beta_2 \sin{\theta_2},& \beta_2 \cos{\theta_2}+ \beta_3 \sin{\theta_3},& \beta_3 \cos{\theta_3} \end{bmatrix}^T.
\end{equation}
When sites $A_1$ and $B_1$ are connected (the initial condition $\theta_{1,i}=\pi/2$, i.e.\ $\sin\theta_1 = 1$) and site $A_2$ is isolated, the $H^{(1)}$ defect eigenstates are $\ket{\Psi_{\varepsilon=0}} = \ket{A_2}$ and $\ket{\Psi_\pm} = (\ket{A_1}\pm\ket{B_1})/\sqrt{2}$. The initial form of the $\ket{B}$ component is $\ket{B_i} = \begin{bmatrix} 1, & 0, & 0 \end{bmatrix}$ and the final form is $\ket{B_f} = \begin{bmatrix} 0, & 0, & -1 \end{bmatrix}$. Setting the connections as
\begin{equation}
\theta_{1,i} = \theta_{2,f} = \pi/2, \quad \theta_{2,i} = 0, \quad \theta_{3,f} = \gamma, \quad \theta_{3,i} = \lambda, \quad \theta_{1,f} = \frac{\pi}{2} - \lambda,
\label{eq:connection_evol_1}
\end{equation}
the global state can be written as
\begin{align}
&\ket{\Psi_{\pm,i}} = \frac{1}{\sqrt{2}}\begin{bmatrix}\pm 1,& 1,& 0,& 0,&0,& 0,& 0\end{bmatrix}^T, \notag \\
&\ket{\Psi_{\pm,f}} = \frac{1}{\sqrt{2}} \begin{bmatrix} 0,& 0,& 0,& 0, & 0, &-1, &\mp \cos{\gamma} \end{bmatrix}^T,
\end{align}
which means that if $\gamma = 0$, a phase shift of $\pi$ will occur for $\varepsilon=1$, but not for $\varepsilon=-1$ (see Fig.~\ref{fig:transfer_states}). However when $\gamma = \pi$, $\varepsilon=1$ will generate an antibonding state from an initial bonding state, and a $\pi$ phase shift will occur when $\varepsilon=-1$ (see Fig.~\ref{fig:transfer_states}). The same results for $\ket{B}$ hold when sites $B_1$ and $A_2$ are connected in the defect and $A_1$ is isolated, but since the hopping angles change to
$
\theta_{1,i} = \theta_{2,i} = 0, \quad \theta_{2,f}= \pi/2,
$
the global state becomes
\begin{equation}
\ket{\Psi_{\pm, i}} = \frac{1}{\sqrt{2}} \begin{bmatrix}0,& \pm 1,& 1,& 0,&0,& 0,& 0\end{bmatrix}^T.
\end{equation}
In general, eigenstates of finite energy ($\varepsilon=\pm1$) acquire a dynamical phase factor $e^{\pm iT/\hbar}$ in the final state, whereas for the zero-energy state this phase is absent.

Additionally, because the lattice is bipartite with a sublattice imbalance, there is a zero-energy state supported entirely on sublattice $A$, with all $B$ sites at zero weight. Setting $A_1=1$ and applying the tight-binding eigenvalue equation gives (apart from a normalization constant)
\begin{equation}
\ket{A} = \begin{bmatrix} 1, & - \tan{\theta_1}, & \tan{\theta_1} \tan{\theta_2}, & - \tan{\theta_1} \tan{\theta_2} \tan{\theta_3} \end{bmatrix}^T.
\end{equation}
This means that if the connection evolution is set as (see Fig.~\ref{fig:transfer_states})
\begin{equation}
\theta_{1,i} = \theta_{2,f} = \pi/2, \quad \theta_{2,i} = \theta_{3,f} = 0, \quad \theta_{3,i} = \gamma, \quad \theta_{1,f} = \pi/2 - \gamma
\end{equation}
the normalized state will evolve from $\ket{\Psi_i} = \ket{A_2}$ to $\ket{\Psi_f} = -\ket{A_3}$, acquiring a $\pi$‑phase. The same will occur if $\theta_{3,f} = \pi$ (see Fig.~\ref{fig:transfer_states}). 

This reasoning applies recursively to the chain $H^{(n)}$ obtained by squaring $H^{(n-1)}$ (see Fig.~\ref{fig:recursive_squaring}). One can therefore construct a protocol that transfers a large number of defect states at once, inheriting the protection of the sine--cosine chain.

\section{Robustness to disorder and level crossing}\label{app:robustness}

This appendix analyzes the robustness of the transfer protocol against disorder and level crossing. The central point is that, without a minimum gap between the Hamiltonian eigenvalues, weight leaks between eigenstates during the evolution, degrading the transfer fidelity. We first show this numerically, then reduce the near-degenerate dynamics to an effective two-level model.

\begin{figure}[ht!]
	\centering  
	\includegraphics[
	width=0.9\textwidth]{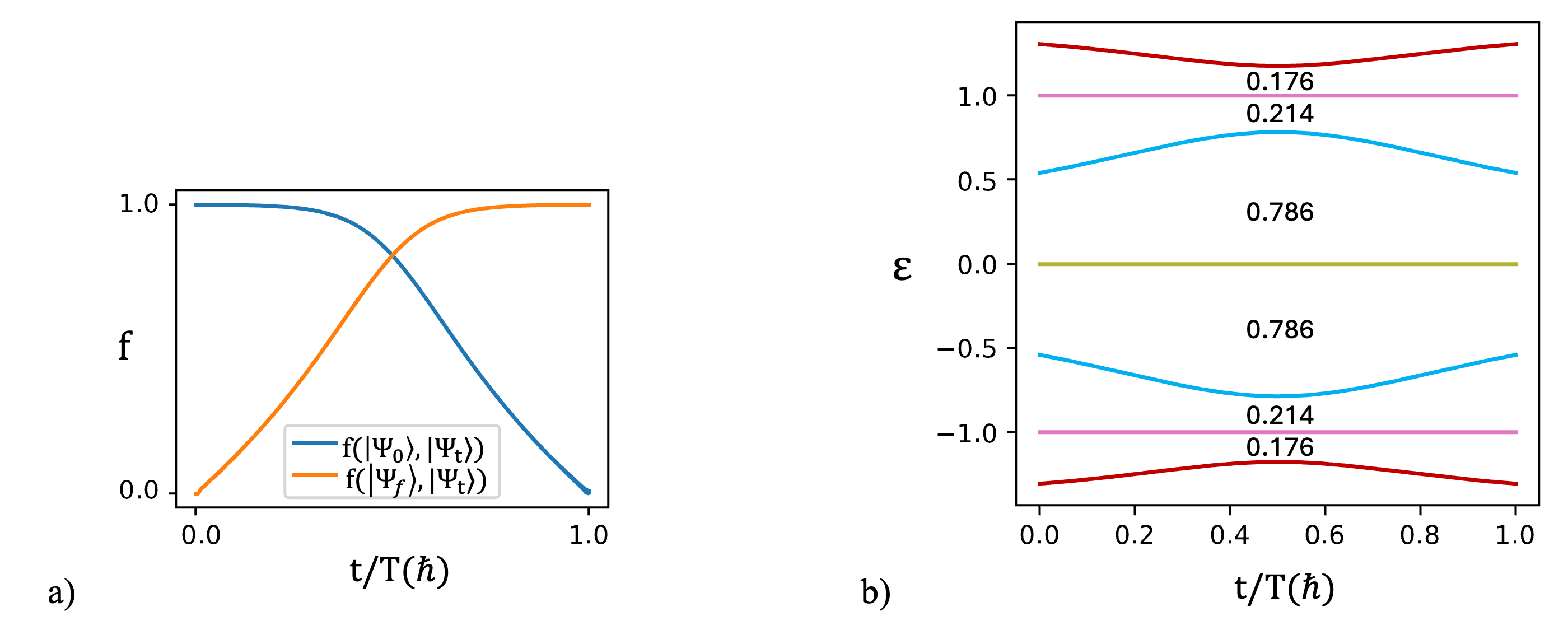}
	\caption{Evolution of $\Psi_0$ under $H^{(1)}$ without disorder using the hopping values in Eq.~\ref{eq:connection_evol_1}. (a) Blue line shows the fidelity evolution compared to the initial state $|\Psi_0\rangle$,while the  orange line shows the  fidelity compared to the expected final state $|\Psi_f\rangle$. (b) Energy spectrum during complete state transfer, where T represents total transfer time.}
	\label{fig:no_disorder_plots}
\end{figure}

\begin{figure}[ht!]
	\centering  
	\includegraphics[
	width=0.9\textwidth]{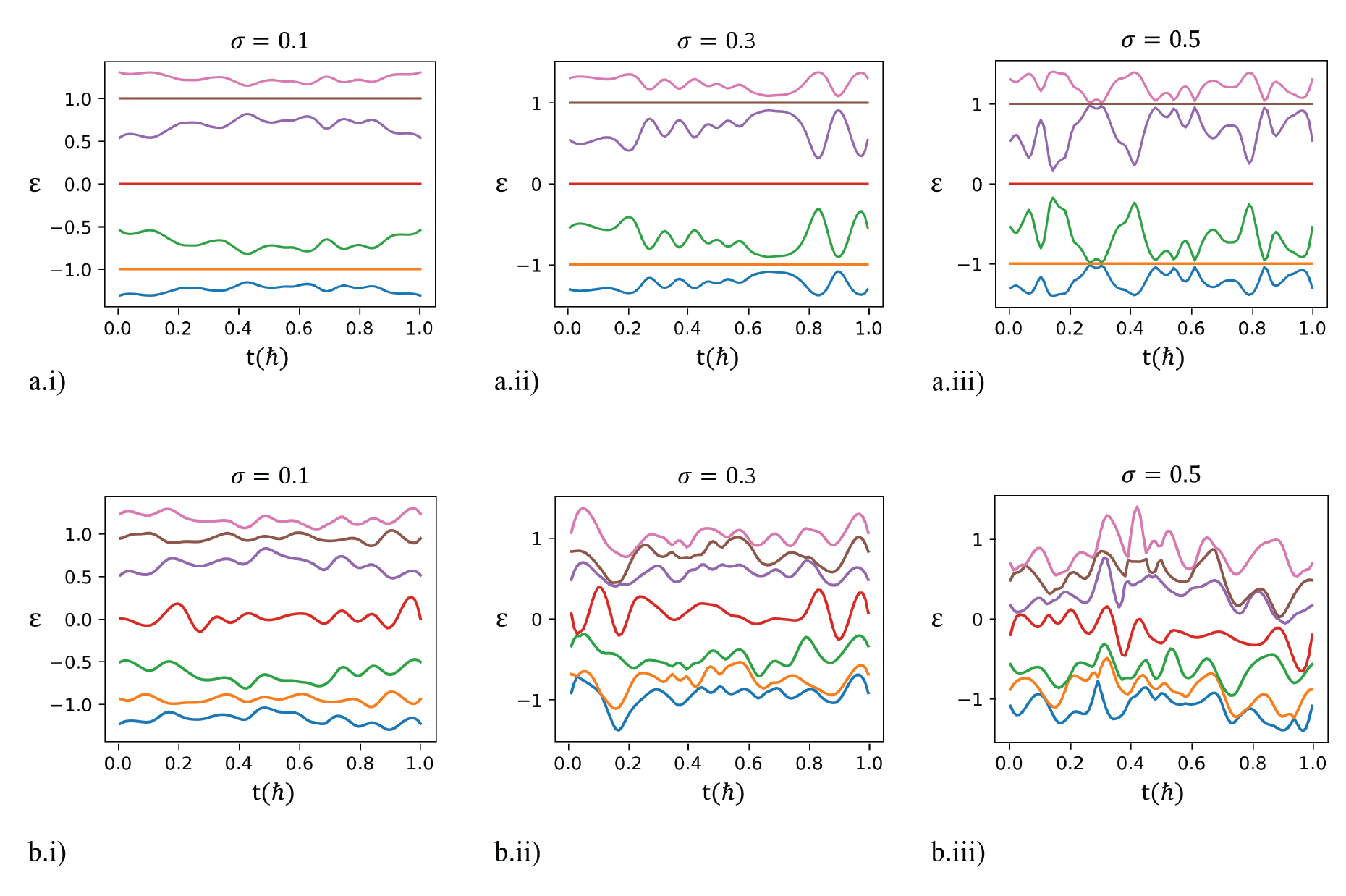}
	\caption{Evolution of the energy spectrum under increasing levels of Gaussian disorder applied on the $\theta$ parameters: a.i) $G(0, \sigma = 0.1, \theta)$, a.ii) $G(0, \sigma = 0.3, \theta)$, and a.iii) $G(0, \sigma = 0.5, \theta)$.  As disorder grows, the levels move closer together and eventually cross, leading to stronger mixing between states. b.i), b.ii), and b.iii): same disorder distribution as in a) but applied to on-site energies.}
	\label{fig:disorder_increase}
\end{figure}

To keep the adiabatic condition, the Hamiltonian's energy levels must stay separated by a finite gap. Without disorder, for the transfer of Fig.~\ref{fig:transfer_states} with $\gamma=0$, the initial state $\ket{\Psi_0} = (\ket{1} + \ket{2})/\sqrt{2}$ evolves to the final state $\ket{\Psi_f} = -(\ket{6} + \ket{7})/\sqrt{2}$, and the fidelity between initial and final states transitions smoothly from $F=1$ to $F=0$ as shown in Fig.~\ref{fig:no_disorder_plots}(a),
while the energy levels remain well separated throughout the evolution, as expected for a Matryoshka-type structure. In the absence of  perturbations, the average fidelity stays close to one, and the minimum level separations during the transfer are 0.786, 0.214, and 0.176, as shown in Fig.~ \ref{fig:no_disorder_plots}(b). Introducing Gaussian disorder in the $\theta$ parameters brings the levels closer together until they cross [see Fig.~\ref{fig:disorder_increase}(a)]. The effect is stronger for on-site disorder, where the crossings occur earlier [see Fig.~\ref{fig:disorder_increase}(b)].
Figure \ref{fig:disorder_increase} also shows that, for disorder in the $\theta$ parameters, the fidelity is expected to drop faster  once $\sigma > 0.3$,  due to the emergence of level crossings that involve the edge states.

To analyze the mechanism behind this fidelity loss, we focus on two states $\ket{\epsilon_i (t)}$ and $\ket{\epsilon_j (t)}$ whose energies $\epsilon_i (t)$ and $\epsilon_j (t)$ intersect at some instant, neglecting weight transfer into the remaining levels (assumed well separated in energy). For a general time-dependent Hamiltonian with a time-dependent eigenbasis $\{ \ket{\epsilon_i (t)}\}$, a state can be expanded as
\begin{equation}
\ket{\Psi (t)} = \sum_j \alpha_j (t) \ket{\epsilon_j (t)}.
\end{equation}
So using the time-dependent Schrödinger equation (with $\hbar=1$), one has
\begin{equation}
i H(t) \ket{\Psi (t)} = \frac{d}{dt} \ket{\Psi (t)} = \sum_j \frac{d\alpha_j (t) }{dt}  \ket{\epsilon_j (t)} + \sum_j \alpha_j (t) \frac{d }{dt}  \ket{\epsilon_j (t)}.
\end{equation}
The projection of this equation into $\ket{\epsilon_i (t)}$ leads to
\begin{equation}
\frac{d\alpha_i}{dt} = -i \epsilon_i(t) \alpha_i (t) - \sum_j \alpha_j \bra{\epsilon_i (t)} \frac{d}{dt} \ket{\epsilon_i (t)}.
\end{equation}
The system can be written as the diagonal Hamiltonian $H_d$ plus the projection matrix $D_{ij} = \bra{\epsilon_i} \frac{d}{dt} \ket{\epsilon_j}$ associated with the rotation of the eigenbasis,
\begin{equation}
\frac{d\Psi}{dt} = (-i H_d + D) \Psi,
\end{equation}
where
\begin{equation}
\Psi = \begin{bmatrix} \alpha_1(t)\\ \alpha_2(t) \\ \vdots \end{bmatrix}; \quad H_d = \begin{bmatrix} \varepsilon_1(t)\\ & \varepsilon_2(t) \\ & & \ddots \end{bmatrix}; \quad D = \begin{bmatrix} -\bra{\varepsilon_1}\frac{d}{dt}\ket{\varepsilon_1} & -\bra{\varepsilon_1}\frac{d}{dt}\ket{\varepsilon_2} & \hdots\\ -\bra{\varepsilon_2}\frac{d}{dt}\ket{\varepsilon_1} & \\ \vdots & & \ddots \end{bmatrix}.
\end{equation}

Under adiabatic conditions, $\Delta \epsilon_{ij} = \epsilon_i-\epsilon_j \gg D_{ij}$, the matrix $D$ acts as a perturbation and the evolution is governed by $H_d$, so $|\alpha_i|$ stays approximately constant. When the two levels cross, however, $\Delta \epsilon_{ij} = 0$ and the projection term dominates, producing an exchange of weight between states $i$ and $j$.

The near-degenerate coupling can be reduced to a two-level toy model \cite{Gouveia2016}: the state evolution is a coherent oscillation of an effective qubit, representable on a Bloch sphere and driven by a rotation Hamiltonian
\begin{equation}
\boldsymbol{H} = \frac{1}{2} \boldsymbol{n}\cdot \boldsymbol{\sigma},
\label{eq:rotatingH}
\end{equation}
with $\bold{n} = (n_x, n_y, n_z)$ as the unit vector, which dictates the rotation axis of the state vector. To show this, we use the density matrix of a qubit 
\begin{equation}
\rho(t) = \frac{1}{2} \left( \mathbb{I} + \boldsymbol{r}(t) \cdot \boldsymbol{\sigma} \right),
\end{equation}
where $\boldsymbol{r}(t)$ is the Bloch vector representing the qubit's state at time $t$. From the relation
$
\langle \sigma_i \rangle = \operatorname{Tr}(\rho(t) \sigma_i) = r_i(t),
$
we find
\begin{equation}
\boldsymbol{r}(t) = \langle \Psi(t)| \boldsymbol{\sigma} | \Psi(t) \rangle.
\label{eq:expectsigma}
\end{equation}
Differentiating with respect to time yields
\begin{align}
\frac{d}{dt} \boldsymbol{r}(t) &= \frac{d}{dt} \langle \Psi(t)| \boldsymbol{\sigma} | \Psi(t) \rangle = \langle \Psi'(t)| \boldsymbol{\sigma} | \Psi(t) \rangle + \langle \Psi(t)| \boldsymbol{\sigma} | \Psi'(t) \rangle = \notag \\
&= i \langle \Psi(t)| \boldsymbol{H}(t) \boldsymbol{\sigma} | \Psi(t) \rangle - i \langle \Psi(t)| \boldsymbol{\sigma} \boldsymbol{H}(t) | \Psi(t) \rangle = i \langle \Psi(t)|  \left[  \boldsymbol{H}(t), \boldsymbol{\sigma}  \right] | \Psi(t) \rangle.
\end{align}
The commutator is evaluated using the Hamiltonian of Eq.~\eqref{eq:rotatingH} and the Pauli-matrix relation $[\sigma_i, \sigma_j] = 2i \, \varepsilon_{ijk} \sigma_k$, where $\varepsilon_{ijk}$ is the Levi-Civita symbol,
\begin{equation}
\left[  \boldsymbol{H}(t), \boldsymbol{\sigma}  \right] = \frac{1}{2} \left[ \mathbf{n}(t) \cdot \boldsymbol{\sigma}, \boldsymbol{\sigma} \right] = i (\boldsymbol{n}(t) \times \boldsymbol{\sigma}).
\end{equation}
Using Eq.~\eqref{eq:expectsigma}, the evolution of $\boldsymbol{r}(t)$ is precession around the axis $\boldsymbol{n}(t)$:
\begin{equation}
\frac{d}{dt} \boldsymbol{r}(t) = - \langle \Psi(t)| \boldsymbol{n}(t) \times \boldsymbol{\sigma} | \Psi(t) \rangle = \boldsymbol{n}(t) \times \boldsymbol{r}(t).
\end{equation}

\begin{figure}[ht!]
 \centering  
 \includegraphics[
width=0.9\textwidth]{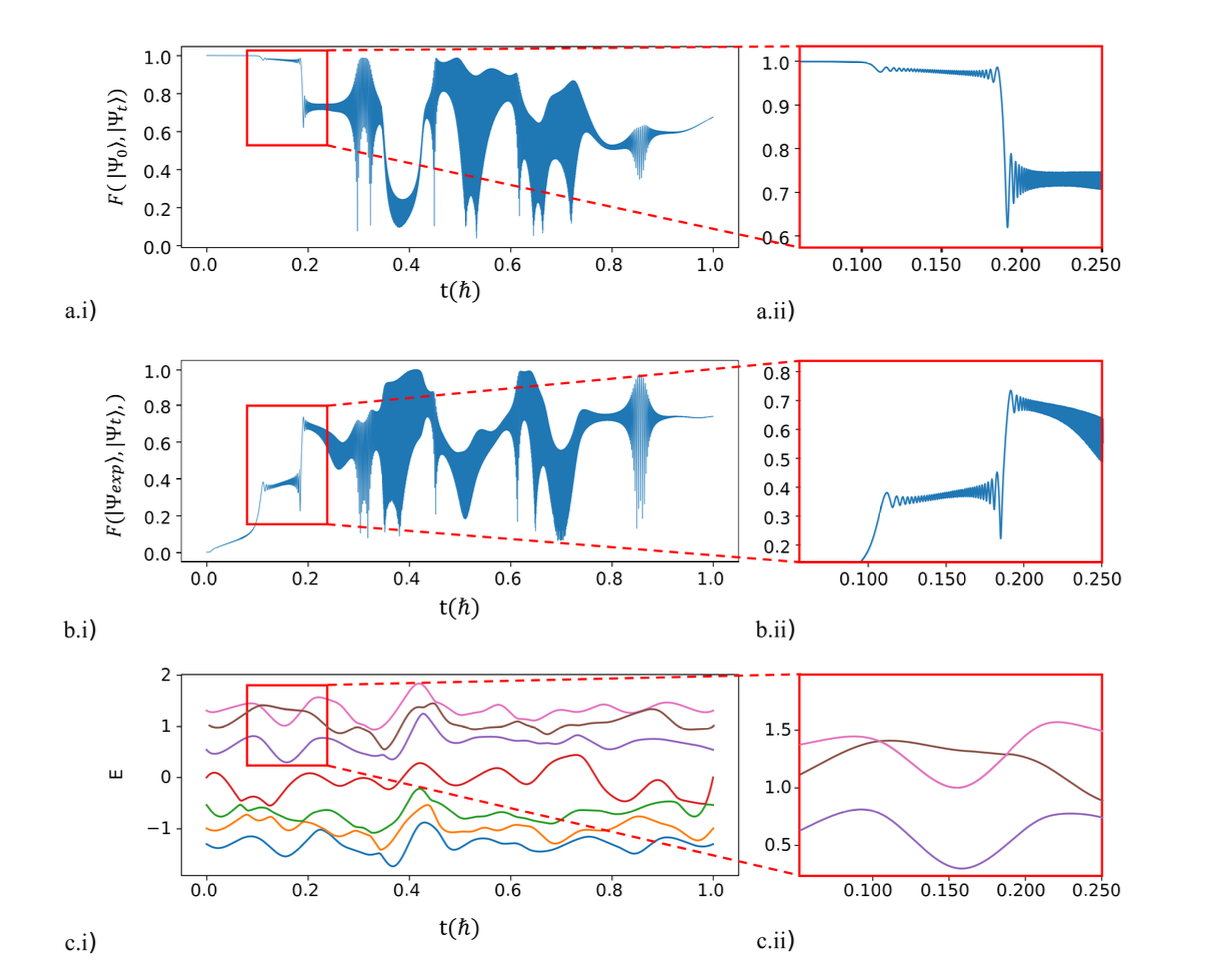}
  \caption{ (a,b) Fidelity evolution of a quantum state with energy initially at  $\varepsilon=1$ of a disordered Matryoshka chain system, with connections defined by Eq.~\ref{eq:connection_evol_1}. (c) Time evolution of the energy spectrum of the chain. Plots show  (a.i) the fidelity relative to the initial state, (b.i) fidelity relative to the expected final state, and (c.i) the time evolution of  energy spectrum. The corresponding insets (a.ii, b.ii, and c.ii) highlight critical regions: fidelity loss during precession near degeneracy points and level crossings that correspond to the observed sharp shifts in fidelity.}
  \label{fig:on_site_disorder_035}
\end{figure}

When the two levels of the effective model cross, a degeneracy occurs. Its eigenvalues are $\varepsilon_{\pm} = \pm \frac{1}{2} |\boldsymbol{n}(t)|$, so the gap is $\Delta \varepsilon = |\boldsymbol{n}(t)| \to 0$ as $\boldsymbol{n}(t) \to 0$. The vector $\boldsymbol{n}(t)$ sets the rotation axis of the Bloch vector: as $\boldsymbol{n}(t) \to 0$ the state vector no longer tracks a well-defined axis and accumulates an angular displacement on the Bloch sphere. When $\boldsymbol{n}(t)$ regrows, the state must rapidly re-align with the new axis, and the fidelity drops abruptly. This weight transfer is the direct consequence of the vanishing gap: the eigenstates mix and the simple precession around a fixed axis breaks down.

As a concrete example, we apply Gaussian disorder with $\sigma = 0.35$ to the on-site energies of the Matryoshka chain. Figures \ref{fig:on_site_disorder_035}(a) and (b) show the sharp fidelity drop at the instant at which two spectral levels cross in Fig.~\ref{fig:on_site_disorder_035}(c); the precession is visible as oscillations in the fidelity, reflecting the exchange of weight between the two higher-energy eigenstates.

\section{Multi-qubit and qudit memory}\label{app:qudit}

In Sec.~\ref{sec:memory} a single qubit was transferred into a Matryoshka-chain memory. This appendix extends the protocol to qudit memory, where higher-dimensional quantum information is stored in the edge states of the same topological framework. The key advantage is that a high-dimensional qudit or, equivalently, several qubits, can be encoded in the edge states of one physical chain, reducing the hardware overhead while retaining the partial topological protection.

The method of Sec.~\ref{sec:memory} can be applied to transfer a qudit with, for example, four components (equivalently, a two-qubit state) into the edge states.
\begin{figure}[ht!]
 \centering  
 \includegraphics[
width=0.7\textwidth]{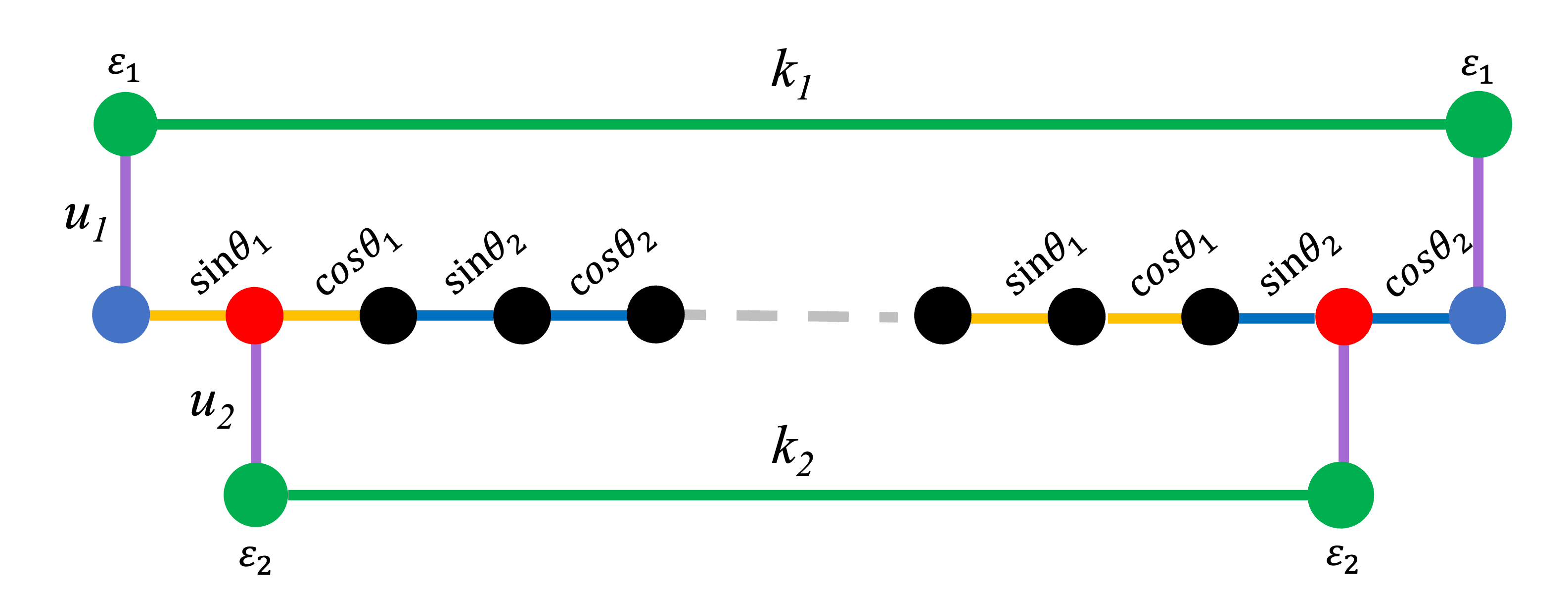}
  \caption{Memory transfer model using a 4-component qudit corresponding to 2 qubits connected to a Matryoshka chain of type $P=2$, with sine‑cosine coupling strengths between memory sites (blue and red circles). Each pair of qudit sites has internal hopping amplitudes $k_1$ and $k_2$ (green lines) and is connected to the memory with hoppings $u_1$ and $u_2$ (purple lines).}
  \label{fig:qudit_memory}
\end{figure}
One approach is to shift the energies of the qudit sites as in Fig.~\ref{fig:qudit_memory} and weakly couple the qudit to the chain so that its components land on edge states of matching energy. This method works but requires long transfer times and precise energy tuning. A more efficient alternative is to extend the protocol of Sec.~\ref{sec:gates}: rapidly transfer the higher-dimensional qudit directly into a Matryoshka $P=2$ chain, choosing the hopping configuration so that specific edge states are selected, as in Fig.~\ref{fig:matryoska_memory}. The transfer is applied separately to each qudit component.

\begin{figure}[ht!]
 \centering  
 \includegraphics[
width=0.7\textwidth]{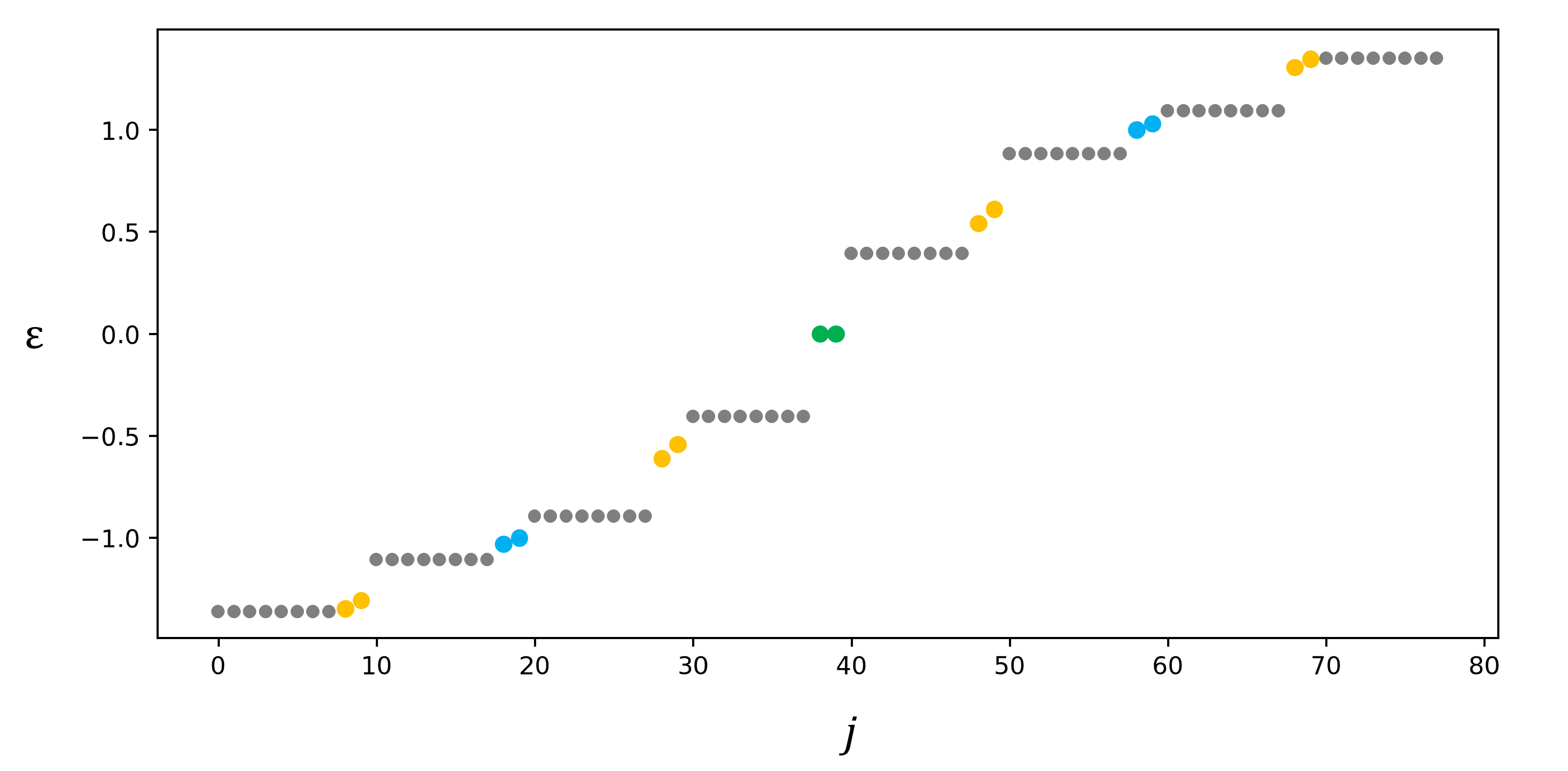}
  \caption{Energy spectrum $\varepsilon$ as a function of the level index $j$ for a Matryoshka-like chain of type $P=2$ with connection angles $\theta_1 = \pi/6$, $\theta_2 = \pi/4$, $\theta_3 = \pi/6$, and $\theta_4 = 0$. The system contains 80 sites, resulting in left and right defects of 7 sites each. The colored points (yellow, blue, and green) correspond to the localized defect states, following the same color scheme as in Fig.~3(c).}
  \label{fig:qudit_spectrum}
\end{figure}

The same construction works for sine--cosine chains that deviate from the exact Matryoshka form. For example, a $P=2$ chain with $\theta_1 = \pi/6$, $\theta_2 = \pi/4$, $\theta_3 = \pi/6$, $\theta_4 = 0$ and 80 sites has a left defect of 7 sites and a right defect of 7 sites, for a total of 14 edge states; the memory capacity is therefore $d=14$. The spectrum for a chain with 80 sites is shown in Fig.~\ref{fig:qudit_spectrum}, with the defect levels in yellow, blue, and green, following the color scheme of Fig.~\ref{fig:bands}(c).

\begin{figure}[ht!]
 \centering  
 \includegraphics[
width=0.9\textwidth]{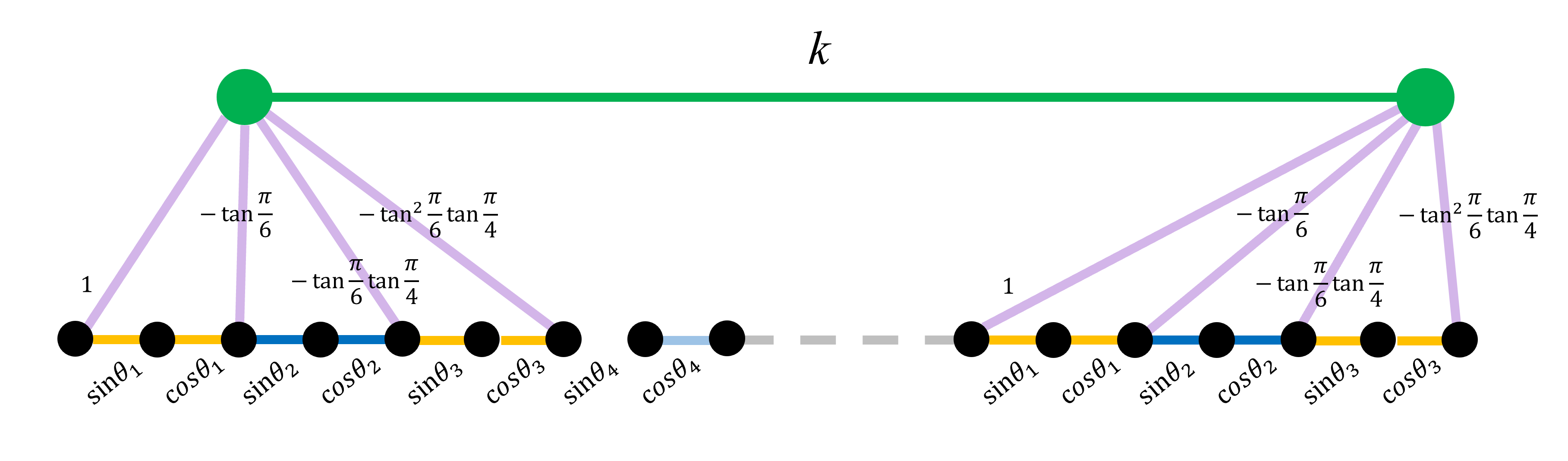}
  \caption{Controlled localization and transfer of a qubit (green sites) to defect states of the quantum memory using a Matryoshka-type chain with connection angles $\theta_1 = \pi/6$, $\theta_2 = \pi/4$, $\theta_3 = \pi/6$, and $\theta_4 = 0$ and 80 sites. The transfer protocol ensures adiabatic evolution.}
  \label{fig:qudit_transfer}
\end{figure}

Diagonalizing the left defect Hamiltonian yields, for example, the zero-energy defect eigenstate (up to normalization),
\begin{equation}
\ket{\psi_{\varepsilon=0}} \propto \begin{bmatrix} 1, & 0, & -\tan(\pi/6), & 0, & -\tan(\pi/6)\tan(\pi/4), & 0, & -\tan^2(\pi/6)\tan(\pi/4) \end{bmatrix}^T.
\end{equation}
These weights (scaled by a constant $u$) are used as the hopping parameters that transfer a qubit into this memory state, as illustrated in Fig.~\ref{fig:qudit_transfer}.
On the right side of the chain the corresponding defect state has approximately the same form.

The qudit (or multi-qubit) memory capacity grows exponentially with the order $P$ of the Matryoshka chain, so a single physical device of order $P$ can store a qudit of dimension up to $2^{P+2}-2$, or equivalently, $(2^{P+1}-1)$ independent qubits, at fixed physical length.

\bibliographystyle{IEEEtran}
\bibliography{references.bib}

\end{document}